\documentclass{article}

\usepackage{arxiv}

\usepackage[utf8]{inputenc} 
\usepackage[T1]{fontenc}    
\usepackage{hyperref}      
\usepackage{url}            
\usepackage{booktabs}       
\usepackage{amsfonts}       
\usepackage{nicefrac}       
\usepackage{lipsum}
\usepackage{graphicx}
\usepackage[style=vancouver]{biblatex}
\usepackage{caption}
\usepackage{subcaption}
\usepackage{textcomp}
\usepackage{gensymb}
\usepackage{xcolor}
\usepackage{soul}
\usepackage{lineno}
\usepackage{setspace}
\usepackage{amsmath}
\usepackage{bbm}

\graphicspath{{images/}}

\title{Comparison of indicators to evaluate the performance of climate models}

\author{
 Mario J. Gómez\\
  Sistema de Estudios de Posgrado-Escuela de Estadística\\
  Universidad de Costa Rica\\
  San José, Costa Rica \\
  \texttt{mariojavier.gomez@ucr.ac.cr} \\
   \And
 Luis A. Barboza \\
 Centro de Investigación en Matemática Pura y Aplicada-Escuela de Matemática\\
  Universidad of Costa Rica\\
  San José, Costa Rica \\
  \texttt{luisalberto.barboza@ucr.ac.cr} \\
  \And
   Hugo G. Hidalgo\\
  Centro de Investigaciones Geof\'isicas and Escuela de F\'isica\\
  Universidad de Costa Rica\\
  San José, Costa Rica \\
  \texttt{hugo.hidalgo@ucr.ac.cr} \\
  \And
   Eric J. Alfaro \\
  Centro de Investigaciones Geof\'isicas, Escuela de F\'isica and Centro de Investigaci\'on en Ciencias del Mar y Limnolog\'ia\\
  Universidad de Costa Rica\\
  San José, Costa Rica \\
  \texttt{erick.alfaro@ucr.ac.cr} \\
}

\begin{document}
\maketitle
\onehalfspacing
\begin{abstract}
The evaluation of climate models is a crucial step in climate studies. It consists of quantifying the resemblance of model outputs to reference data to identify models with superior capacity to replicate specific climate variables. Clearly, the choice of the evaluation indicator significantly impacts the results, underscoring the importance of selecting an indicator that properly captures the characteristics of a "good model". This study examines the behavior of six indicators, considering spatial correlation, distribution mean, variance, and shape. A new multi-component measure was selected based on these criteria to assess the performance of 48 CMIP6 models in reproducing the annual seasonal cycle of precipitation, temperature, and teleconnection patterns in Central America. The top six models were determined using multi-criteria methods. It was found that even the best model reproduces one derived climatic variable poorly in this region. The proposed measure and selection method can contribute to enhancing the accuracy of climatological research based on climate models.
\end{abstract}

\keywords{Climate models evaluation \and GCMs \and CMIP6 \and Central America}

\section{Introduction}

Climate models are complex mathematical representations of the components of the climate system \cite{GFDL}. Classified and ordered by their level of complexity, the General Circulation Models (GCMs) are located at the far end of the line. They are characterized by their global dominion and a large range of oceanic and atmospheric processes \cite{Katzav-2015}. Considering that these models are developed by researchers in different laboratories across the globe under different methodologies and assumptions (including different physics and initializations), is not a surprise to find divergences in their projections\cite{Nguyen-2017}. The evaluation of these climate models consists on the analysis of the differences between these projections and a set of observed or reanalysis data \cite{IPCC-2014}, in order to identify the ones with the best capacity to reproduce a certain climatic characteristic in a specific spatial and temporal frame. It is evident that this performance is affected by the particular variable to be measured, the geographical area and the selected period, the data used as reference and finally the selected measure to quantify similarity \cite{Pincus-2008}. Despite the plethora of available statistics to accomplish this evaluation task \cite{Raju-2014}, there is not a standard method, because its selection will depend on the measurement purpose \cite{Gleckler-2008}, which makes the exploration of new techniques still necessary \cite{Knutti-2010}. A flaw in this exploration is the proposal of new measures without the assurance that they are capturing properly the characteristic of a ``good model''. Of course, the definition of a ``good model'' is totally subjective, and can vary according to the goals settled, but at least, it must be specified for each evaluation method. In this study, the criteria proposed by Taylor \cite*{Taylor-2001} is used: a good model generates data that presents a high linear correlation with the reference set, besides similar variation and mean. Additional to these criteria, the similarity between the distributions shapes of the model output and the observed or reanalysis data is also contemplated. The importance of this last feature was pointed by Demirel et al. \cite{Demirel-2018}.

The model outputs used in this assessments are of a spatio-temporal nature. This means that every single data point is associated to a spatio-temporal location \cite{Tan-2017}. According to Wikle et al.\cite*{Wikle-2019} this kind of data can be seen as a sequence of static states or ``pictures'' (spatial perspective) or as a collection of time series (temporal perspective). This perspective is intrinsically related to the evaluation measures, which can be classified accordingly.

This research employs both spatial and temporal perspective methods, with the temporal approach relying on Functional Data Analysis (FDA) \cite{Ramsay-1982} - a statistical methodology that treats each indivisible unit of analysis as a function \cite{Suhaila-2017}.

Our primary purpose is to evaluate the effectiveness of these measures in identifying a good model based on the Taylor criteria. This work also aims to determine which methods are suitable for generating ranking exercises using real climate model data from the Coupled Model Intercomparison Project (CMIP), currently in its sixth phase \cite{Eyring-2016}. The structure of this article is as follows: Firstly, a description is provided for each of the techniques being considered. Next, the behavior of each technique is illustrated. Finally, based on these behaviors, a specific measure is chosen to evaluate 48 GCMs based on their ability to replicate the annual cycles for precipitation, temperature and influential teleconnections in Central America.

\section{Data and Methods}

\subsection{Performance Measures}

A completed and detailed list of measures to evaluate the performance of climatic models is beyond the scope of this work. However, after a thorough examination of various techniques, we have selected the following based on their potential ability to identify correlation and distribution similarities with reference data. 

\subsubsection{Skill Score}

This is one of the most used methods to evaluate model performance. Proposed by Murphy \cite*{Murphy-1988}, it is defined as the relative precision between two forecasts, fixing one of them as reference. Using the Mean Square Error (MSE) as the precision measure and recurring to the sample mean of the observed values as reference, the Skill Score of the model output $\mathbf{y}$ with respect to the observed values $\mathbf{x}$ is

\begin{equation*}
SS(\mathbf{x},\mathbf{y}) = 
\rho_{xy}^2 - 
\Big[\rho_{xy} - \frac{s_{y}}{s_{x}}\Big]^2 - \Big[\frac{\bar{y}-\bar{x}}{s_{x}}\Big]^2.
\end{equation*}

where $\rho_{xy}$ represents the linear correlation between the $\mathbf{x}$ and $\mathbf{y}$ vectors, $s_{x}$ and $s_{y}$ their standard deviations and $\bar{x}$ and $\bar{y}$ their respective means. The resulting values can vary from $-\infty$ to 1, being 1 the perfect score. Hidalgo and Alfaro \cite*{Hidalgo-2012} already applied this method to evaluate  climate change projections in the Eastern Tropical Pacific Seascape and to rank CMIP5 models for reproducing characteristics of the seasonal cycles and El Niño-Southern Oscillation (ENSO) signal in Central America's climate \cite{Hidalgo-2015}.

\subsubsection{Spatial Efficiency}

Demirel et al.\cite{Demirel-2018} proposed the Spatial Efficiency (SPAEF) measure as an improvement of the Kling Gupta Efficiency (KGE) \cite{Gupta-2009,Centella-2020}, which is an attempt to correct some deficiencies of the Skill Score. It is conceived under a multi-objective approach in the calibration of hydrological models. It considers the Euclidean distance of three equally weighted components, as follows:

\begin{equation*}
SPAEF(\mathbf{x},\mathbf{y}) = 
1 - \sqrt{(\alpha-1)^2 + (\beta-1)^2 +
(\gamma-1)^2}
\end{equation*}

where $\alpha = \rho_{xy}$, $\beta = \frac{s_{y}/s_{x}}{\bar{y}/\bar{x}}$ and $\gamma=\sum_{b = 1}^{B} \min(h_{xb},h_{yb})/\sum_{b = 1}^{B}h_{xb}$. The last term represents the intersection of histograms $h_{x}$ and $h_{y}$ comparing their common $B$ bins, according to Swain \& Ballard \cite{Swain-1991}. The range of values for this measure extends from $-\infty$ to 1, with a higher value closer to 1 denoting a stronger agreement with the reference data.

This indicator has been used by Ahmed et al. \cite{Ahmed-2019} in the ranking of GCMs in Pakistan.

\subsubsection{Wasserstein Distance}

The concept of Wasserstein distance (WD) arises from the optimal mass transportation problem and serves as a metric family specifically tailored for analyzing probability distributions \cite{Villani-2009}. It quantifies the cost of moving mass from one location to another. The WD of order $p$ between two probability distributions $P$ and $Q$ in $\mathbbm{R}^d$ is given by \cite{Panaretos-2019}:

\begin{equation*}
WD(P,Q) = \inf\limits_{\substack{X \sim P \ Y \sim Q}}(\mathbbm{E}|X - Y|)^{\frac{1}{p}} \quad p > 0,
\end{equation*}

where $X$ and $Y$ are random variables following distributions $P$ and $Q$, respectively. If $P$ and $Q$ are the empirical distributions of $\mathbf{x}$ and $\mathbf{y}$, then the WD of order $p$ can be expressed as:

\begin{equation*}
WD_{p}(\mathbf{x},\mathbf{y})=\Bigg(
\sum_{i = 1}^{n}|x_{i}-y_{i}|^{p}
\Bigg)^\frac{1}{p}
\end{equation*}

$p=2$ is used in this study. As a distance metric, it takes values within the interval $[0,+\infty)$.

Vissio et al. \cite*{Vissio-2020} applied this distance in the evaluation of global-scale climate models.

\subsubsection{Wasserstein Spatial Efficiency}

As mentioned earlier, the effectiveness of a performance indicator in the context of this study, is subject to its capacity to identify spatial correspondence and differences between distributions: their position, dispersion and shape. The previous measures fail to detect at least one of these attributes.  For example, SPAEF cannot quantify differences in the mean when the histogram intersection is applied to standardized vectors. On the other hand, WD is inadequate for measuring correlation since it does not consider the position of the values.. To address these limitations and provide a more comprehensive evaluation, we propose a novel measure that combines features from both indicators. This is accomplished by replacing the histogram intersection $\gamma$ in the SPAEF formulation by WD=$\phi$, because this distance is defined for not overlapping distributions \cite{Bernton-2019}, which allows to account for the bias. The Wasserstein Spatial Efficiency (WSPAEF) is the defined as:

\begin{equation*}
WSPAEF(\mathbf{x},\mathbf{y}) = 
\sqrt{(\alpha-1)^2 + (\varsigma-1)^2 + \phi^2}
\end{equation*}

\noindent where $\varsigma = s_{y}/s_{x}$, due the fact that Wasserstein distance is applied over the scaled vectors with variance equal to 1. The preference over the original $\beta$ term arises from its instability when the means approach zero. This instability can occur in the evaluation of dry or cold regions where precipitation or temperature values are low \cite{Tang-2021}. Initially, the utilization of Wasserstein distance (WD) emerged as a means to circumvent the cross-correlation issue associated with the bias component \cite{Kling-2012}. However, the incorporation of WD resolves this concern, rendering it irrelevant. The values assigned by this indicator range from 0 to $+\infty$ and can be interpreted as a distance from the reference set.

\subsubsection{Kolmogorov Ranking Statistic}

Harris et al. \cite*{Harris-2020} evaluate the similarity of two spatiotemporal fields using the concept of depth, specifically an adaptation of Tukey depth \cite*{Tukey-1975} to functional data. They proposed the Kolmogorov Depth Statistic (KD), which is effective in the detection of parameter's changes and useful and pertinent for hypothesis testing purposes, but it does not go far enough if the final objective is to rank the distributions, when not only is necessary to test if there is a difference but also to determine the magnitude of this difference. For this reason, as an additional contribution of this article, the KD has been modified to quantify the misalignment between two function groups and to explore the behavior of temporal perspective techniques, resulting in the Kolmogorov Depth Ranking (KDPR) statistic. 

As the KD authors explain, If $P$ is a distribution of a random variable $X \in C[0,1]^{p}$, and $P_{s}$ is the marginal distribution $P$ at location $s \in [0,1]^p$, the integrated Tukey depth of $X=x$ wrt $P$ is

\begin{equation*}
D(x,P)=\int_{[0,1]^{p}} D(x(s),P_{s}) ds
\end{equation*}

where $D(x(s),P_{s})$ is the univariate Tukey depth of $x(s)$ wrt to $P_{s}$, defined as $1-|1-2P(x(s))|$.

Consider two samples, $X \sim P$ and $Y \sim Q$, with estimators $P_{n}$ and $Q_{m}$ respectively. The outlyingness of any curve $x_{k}$ wrt to $P$ can be cuantified as:

\begin{equation*}
\hat{F}_{X}(x_{k}) = 
\frac{1}{n} \sum_{i=1}^{n}
(D(x_{i},P_{n}) - D(x_{k},P_{n}))
\end{equation*}

In the original form, this measure is a standardized rank of curves that follows a discrete uniform distribution \cite{Harris-2020}. However, this characteristic reduces its usefulness for ranking purposes.

In the same way as it was defined for $\hat{F}_{X}$, the outlyingness of any curve $y_{k}$ wrt to $P$ is:

\begin{equation*}
\hat{G}_{Y}(x_{k}) = 
\frac{1}{n} \sum_{j=1}^{n} 
(D(y_{j},P_{n}) - D(x_{k},P_{n}))
\end{equation*}

Due, to the asymmetry property of depth, equivalent measures have to be calculated fixing $Q_{m}$. 

\begin{equation*}
\hat{F}_{X}(y_{k}) = 
\frac{1}{n} \sum_{i=1}^{n}
(D(x_{i},Q_{m}) - D(y_{k},Q_{m}))
\end{equation*}

\begin{equation*}
\hat{G}_{Y}(y_{k}) = 
\frac{1}{n} \sum_{j=1}^{n}
(D(y_{j},Q_{m}) - D(y_{k},Q_{m}))
\end{equation*}

The Kolmogorov distance is utilized to compute the KDPR statistic by summing the differences between the ``intra'' outlyingness measures ($\hat{F}_{X}(x_{k})$, $\hat{G}_{Y}(y_{k})$) and their ``inter'' counterparts ($\hat{F}_{X}(y_{k})$, $\hat{G}_{Y}(x_{k})$):

\begin{equation*}
KDPR(X,Y) = \max \left[
\sum \hat{F}_{X}(x_{k})-\hat{G}_{Y}(x_{k}),
\sum \hat{F}_{X}(y_{k})-\hat{G}_{Y}(y_{k})\right]
\label{kdpr}
\end{equation*}

The same approach can be applied by substituting depth with alternative distances, (Kolmogorov Distance Ranking or KDSR). For instance,  $\hat{F}_{X}(x_{k})$ can be redefined as:

\begin{equation*}
\hat{F}_{X}(x_{k}) = 
\frac{1}{n} \sum_{i=1}^{n}
||x_{i}-x_{k}||
\end{equation*}

The range of both versions extends from 0 to $+\infty$, where 0 represents a perfect match between observed-reanalysis and simulated data.

\subsection{Measures Behavior}

To establish our definition of a ``good model'', a measure should operate in the following manner when comparing model and reference sets: it should exhibit an increase in value as the correlation between the sets diminishes and the discrepancies in means and variations widen. Additionally, greater values are expected if the empirical distributions of the sets are not similar in terms of their forms. While the SS behavior can be analytically described in a trivial manner, this is not the case for the rest of the techniques, therefore, it is necessary to generate synthetic data to show their behavior. The realizations $\mathbf{x}$ and $\mathbf{y}$ of two spatiotemporal fields $X$ and $Y$ were compared under the resulting combinations of the following characteristics: (a) Linear correlation ($\rho = \rho_{xy}$): -0.9, -0.3, 0.3, 0.9. (b) Standard deviation ratio ($\lambda = \frac{s_{y}}{s_{x}}$): 0.1, 0.7, 1.0, 1.3, 1.9. (c) Bias ($\delta = |\bar{x} - \bar{y}|$): 0.0, 1.0, 5.0. To achieve this, preliminary samples $\mathbf{x'}$ and $\mathbf{y'}$ were first obtained, which also come from the aforementioned fields $X$ and $Y$, respectively. These consist of sets of 100 static states (temporal segments) of size $5 \times 5$, generated using an isotropic and stationary covariance model from the Matérn family, according to the equation:

\begin{equation*}
C(r)=\frac{2^{1-\nu}}{\Gamma{\nu}} (\sqrt{2\nu r})^{\nu} K_{\nu} (\sqrt{2\nu r})
\end{equation*}

where $r$ is the distance between two points, $\nu$ is the smoothing parameter ($\nu = 1.5$ in this design), and $K_{\nu}$ is the modified Bessel function of the second kind \cite{RandomFields-2015}. In addition to the established scenarios, two conditions were created: ``Undisturbed'' where the shape of the empirical distribution of the preliminary samples $\mathbf{x'}$ or $\mathbf{y'}$ is not altered, and ``Disturbed'' where a log-normal transformation ($\mu = 1$ and $\sigma = 0.5$) is applied to $\mathbf{x'}$. For each combination, the data composing $\mathbf{x'}$ and $\mathbf{y'}$ were organized into centered vectors $\mathbf{u}$ and $\mathbf{v}$. Using $\mathbf{u}$ as a reference and considering that correlation is the magnitude of the cosine of the angle $\theta$ between two centered vectors \cite{Shevlyakov-2016}, a new vector $\mathbf{w}$ was generated, defined as:

\begin{equation*}
\mathbf{w} = \rho_{uw}\frac{\mathbf{v}_{\parallel \mathbf{u}}}{\|\mathbf{v}_{\parallel \mathbf{u}}\|} + \sqrt{1-\rho_{uw}^2}\frac{\mathbf{v}_{\perp \mathbf{u}}}{\|\mathbf{v}_{\perp \mathbf{u}}\|} 
\end{equation*}

where $\rho_{uw}$ is the desired correlation between the vectors $\mathbf{u}$ and $\mathbf{w}$, $\mathbf{v}_{\parallel \mathbf{u}}$ is the projection of $\mathbf{v}$ onto $\mathbf{u}$, and $\mathbf{v}_{\perp \mathbf{u}} = \mathbf{v} - \mathbf{v}_{\parallel \mathbf{u}}$ is the projection of $\mathbf{v}$ onto the orthogonal complement of $\mathbf{u}$. Since correlation is not affected by linear transformations, the reference vector $\mathbf{x} = \frac{\mathbf{u}}{s_{u}+c}$ and the vector $\mathbf{y} = \frac{\mathbf{w}}{s_{w}\lambda + c + \delta}$ were defined such that $\lambda = \frac{s_{y}}{s_{x}}$, $\delta = |\bar{x} - \bar{y}|$ (as required), and $\rho_{uw}= \rho_{xy}$. The constant $c$ is arbitrary, as the interest lies in the relationships between certain characteristics of the samples rather than their values. In this case, $c = 10$ (because of the issues related to some indicators when the mean is close to 0). Finally, the selected techniques were applied to the pair of vectors $\mathbf{x}$ and $\mathbf{y}$ in each case, considering the entire temporal extension.

\begin{figure}[p]
  \centering
  \includegraphics[scale=0.85]{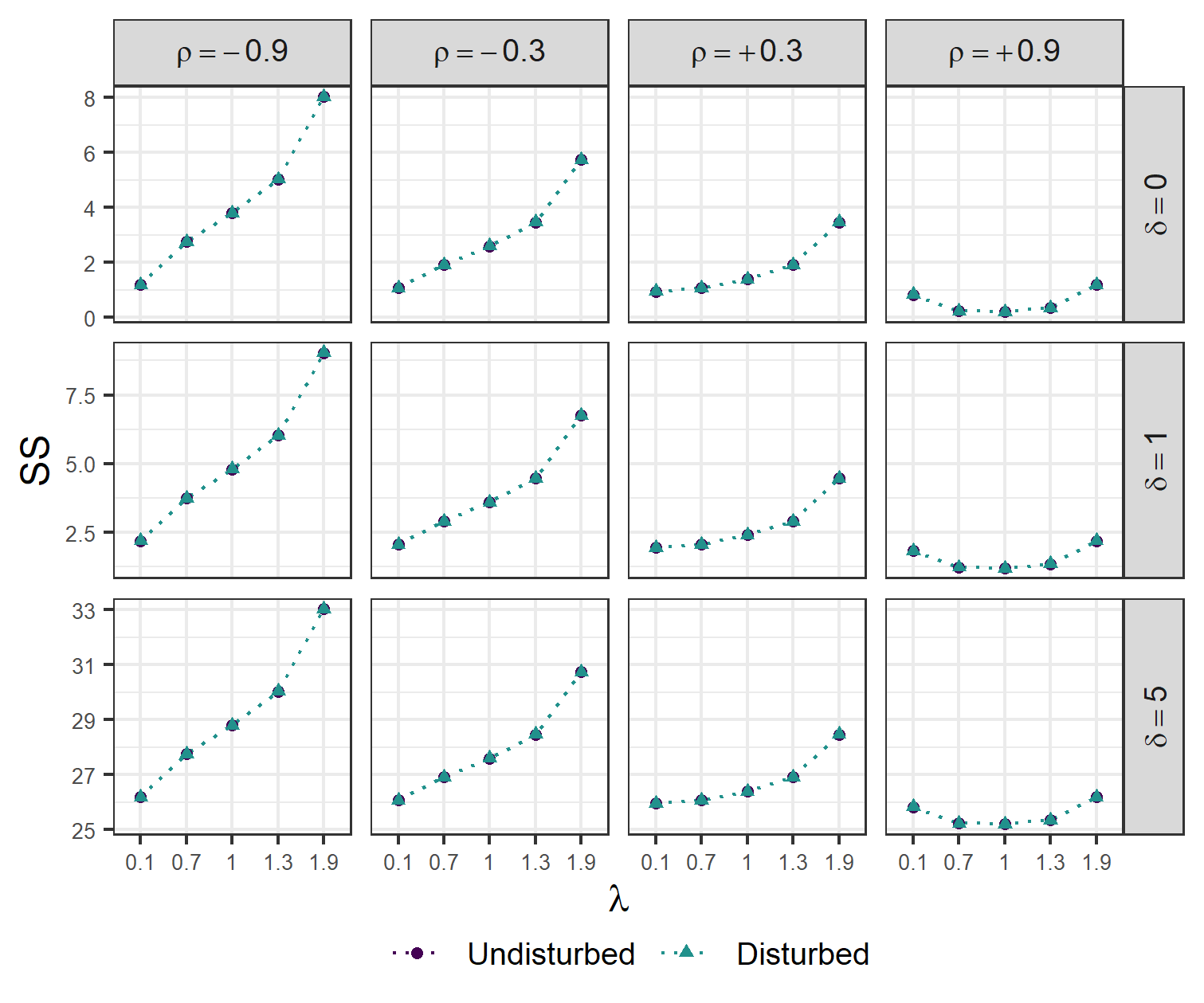}
  \caption{Skill score behavior}
  \caption*{Notes: $\rho$: linear correlation, $\lambda$: variation ratio, $\delta$: bias. The ``Undisturbed'' series refers to normal data while the ``Disturbed'' series refers to a log-normal transformation of this data. Lower values indicates a better model.}
  \label{fig:sscore}
\end{figure}

\begin{figure}[p]
  \centering
  \includegraphics[scale=0.85]{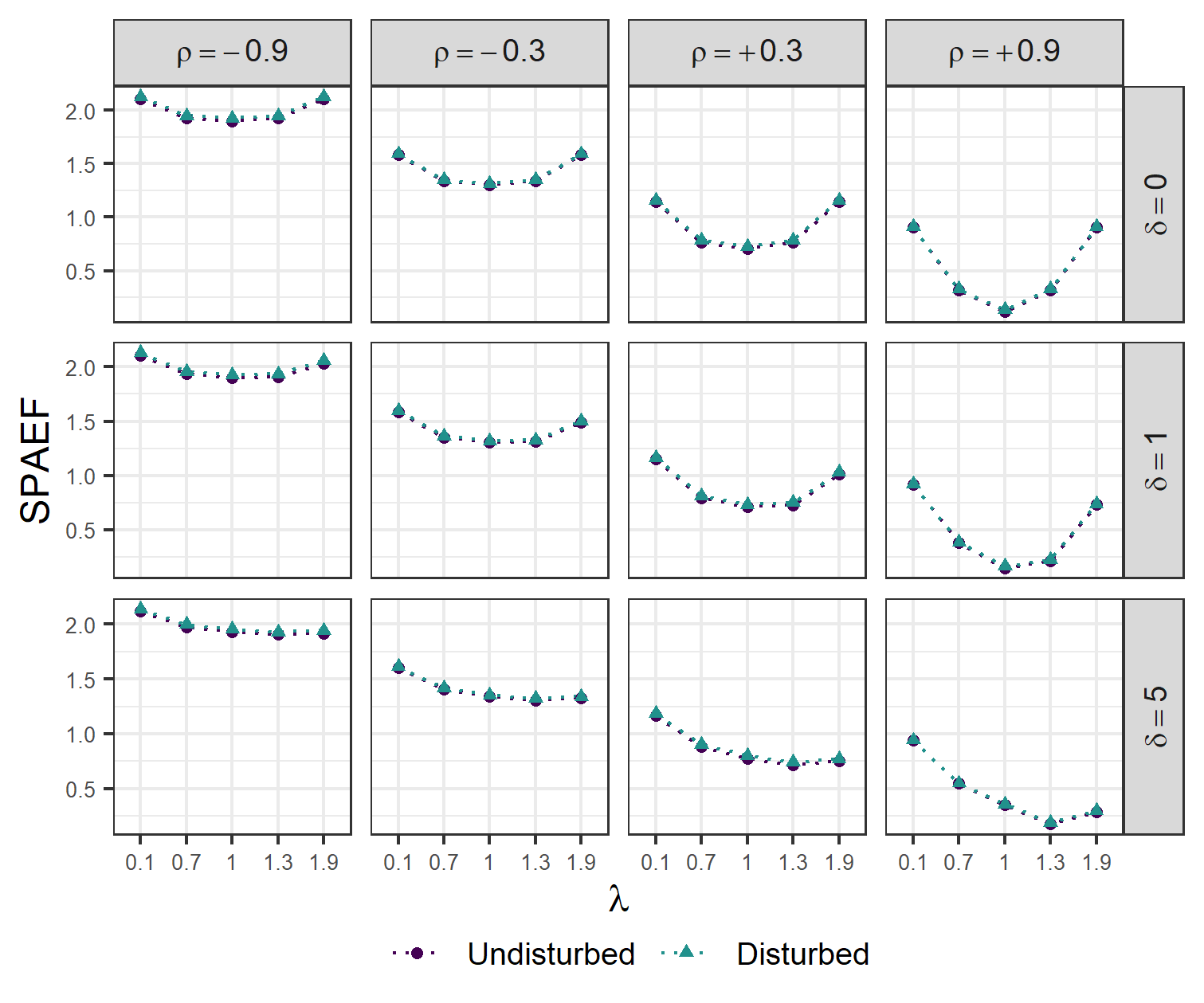}
  \caption{SPAEF behavior}
    \caption*{Notes: $\rho$: linear correlation, $\lambda$: variation ratio, $\delta$: bias. The ``Undisturbed'' series refers to normal data while the ``Disturbed'' series refers to a log-normal transformation of this data. Lower values indicates a better model.}
  \label{fig:spaef}
\end{figure}

\begin{figure}[p]
  \centering
  \includegraphics[scale=0.85]{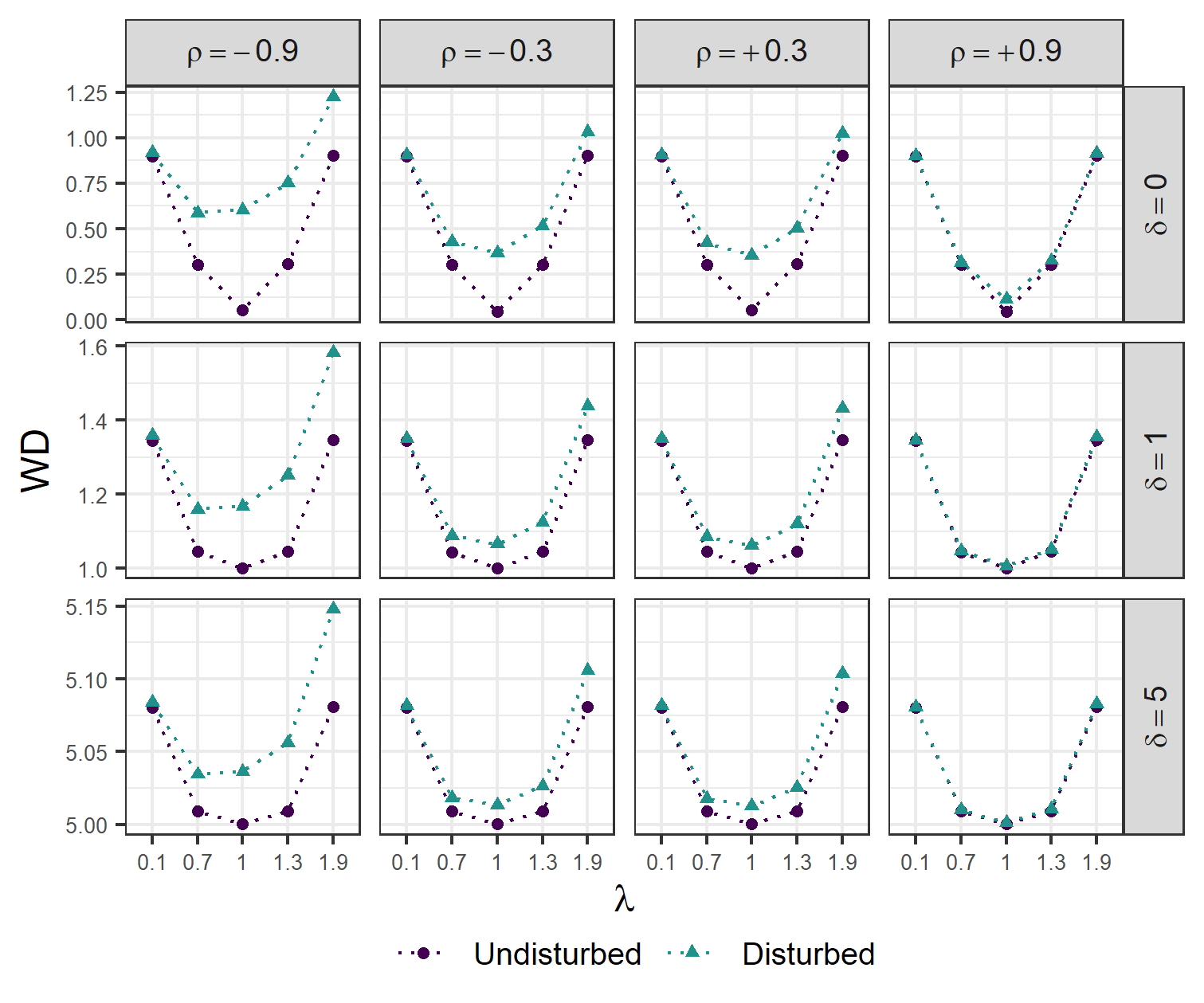}
  \caption{Wasserstsein distance behavior}
    \caption*{Notes: $\rho$: linear correlation, $\lambda$: variation ratio, $\delta$: bias. The ``Undisturbed'' series refers to normal data while the ``Disturbed'' series refers to a log-normal transformation of this data. Lower values indicates a better model.}
  \label{fig:wass}
\end{figure}

\begin{figure}[p]
  \centering
  \includegraphics[scale=0.85]{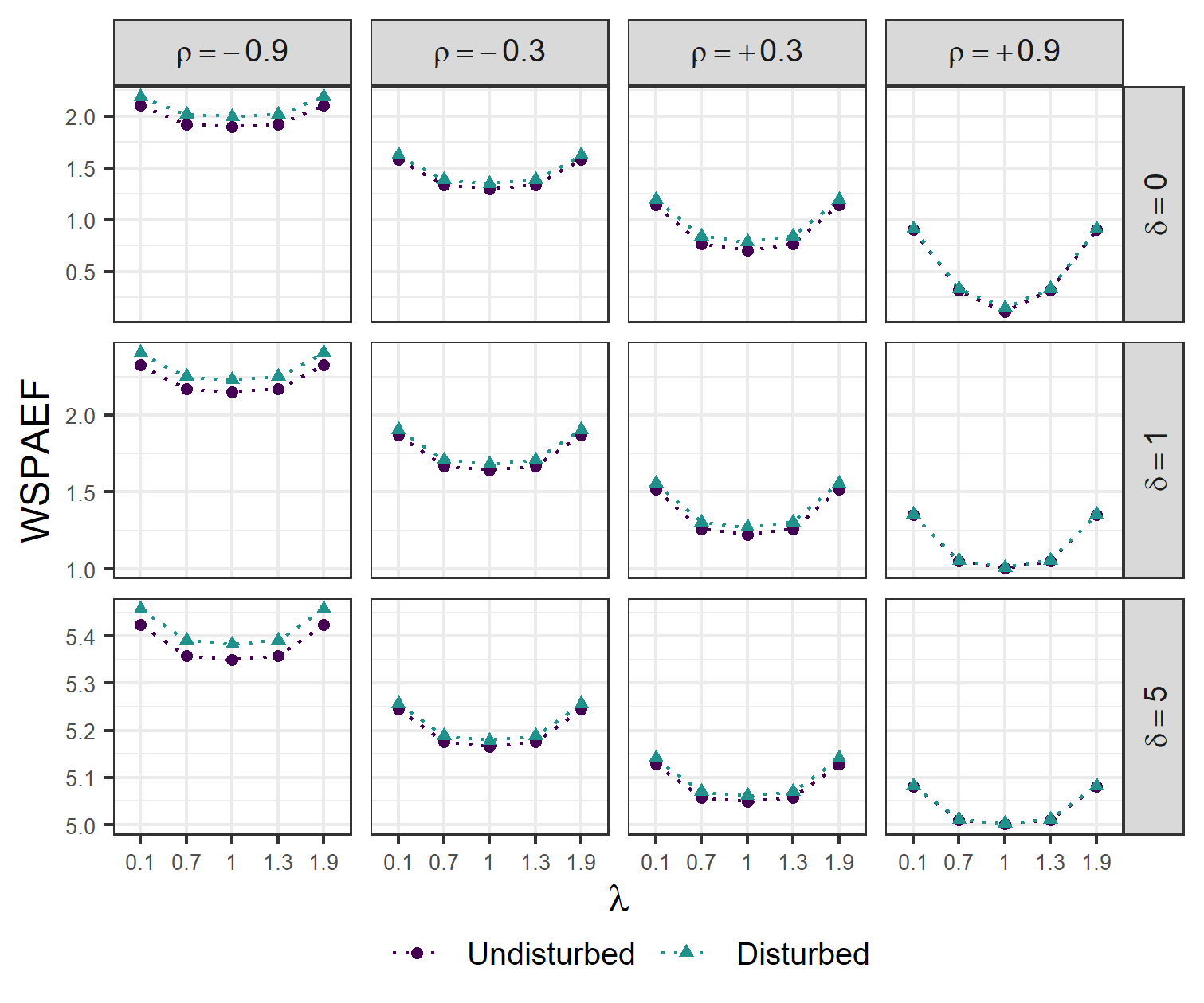}
  \caption{Wasserstein SPAEF}
    \caption*{Notes: $\rho$: linear correlation, $\lambda$: variation ratio, $\delta$: bias. The ``Undisturbed'' series refers to normal data while the ``Disturbed'' series refers to a log-normal transformation of this data. Lower values indicates a better model.}
  \label{fig:spaefws}
\end{figure}

\begin{figure}[p]
  \centering
  \includegraphics[scale=0.85]{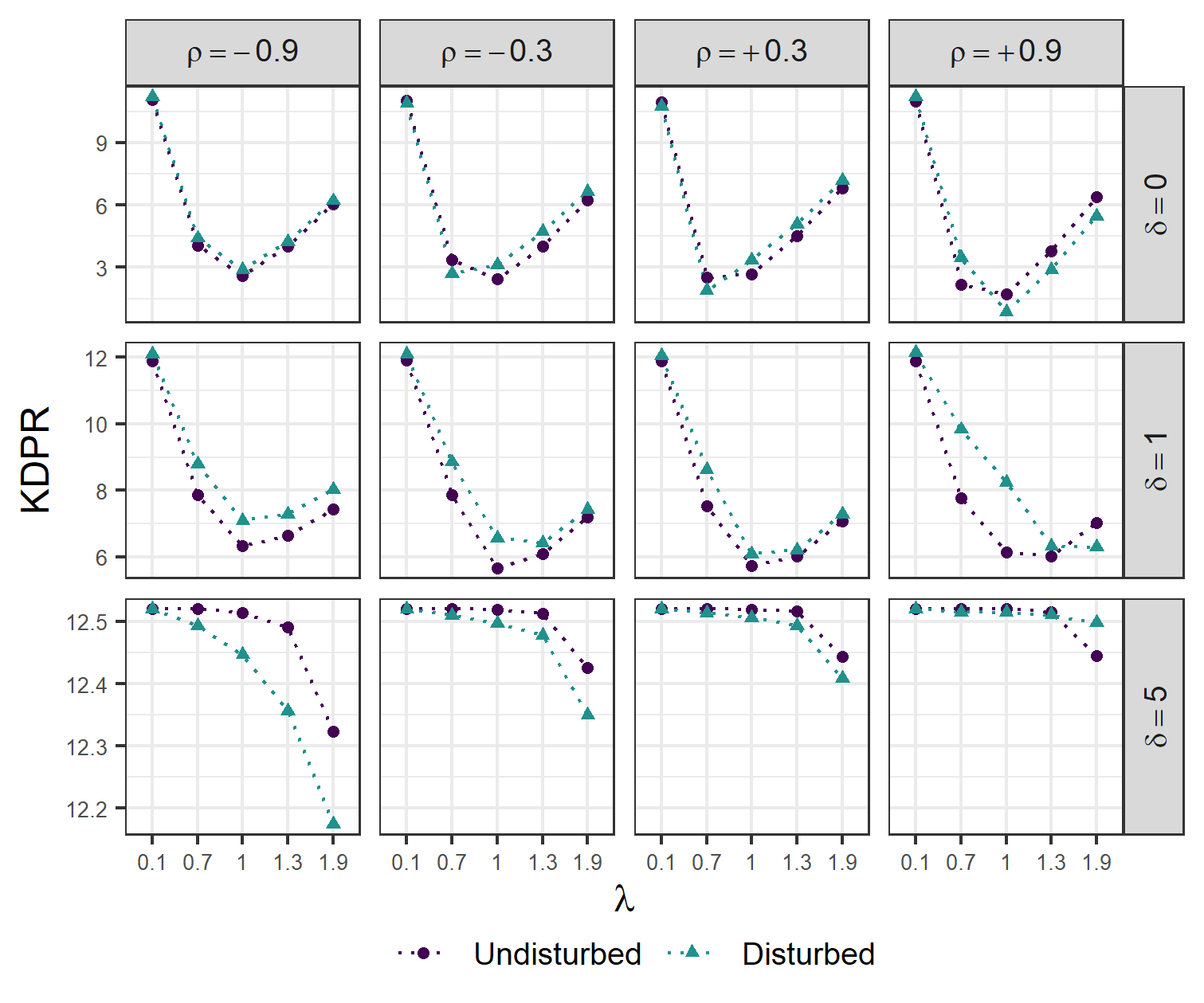}
  \caption{Kolmogorov depth}
    \caption*{Notes: $\rho$: linear correlation, $\lambda$: variation ratio, $\delta$: bias. The ``Undisturbed'' series refers to normal data while the ``Disturbed'' series refers to a log-normal transformation of this data. Lower values indicates a better model.}
  \label{fig:ksp}
\end{figure}

Starting by the SS indicator (Figure \ref{fig:sscore}), as a desirable feature, it shows high sensitivity to bias and correlation, however, only in high positive correlation scenarios the lowest point in each facet correspond to the best model in terms of variability, in any other case, the measure will be oriented to models with lower standard deviations than the reference. The SPAEF  (Figure \ref{fig:spaef}) corrects this last characteristic, but just for models with non discrepancies in the mean. When these discrepancies increase, the SPAEF privileges models with higher variability (relative to the reference). Other problems are the lack of sensitivity to the mean and shape dissemblance. The first issue was pointed out previously, but the second one was expected to be captured with the $\gamma$ component, which is its main feature over the SS or other measures as the Kling Gupta Efficiency. The WD (Figure \ref{fig:wass}) detects effectively the bias, variability (with certain level of asymmetry) and shape but, as mentioned, it does not consider the linear correlation. Adding dimensions solve this last issue, but the resulting behavior is almost identical to the skill score. The WSPAEF (Figure \ref{fig:spaefws}) seems to meet all the requirements. The weight of mean difference and the correlation value is clearly greater that dissimilarities in variation and shape, which can be seem as a desirable feature depending of the application. Finally, the modified Kolmogorov depth (Figure \ref{fig:ksp}) works relative well when $\delta = 0$ but for greater values the behavior change radically until achieving even a concave form. The distance version (Figure \ref{fig:ksp}), leaving aside certain asymmetry, behaves properly when the bias is low. One can said that, in that specific condition, it is the best option to consider if the primary interest is to detect differences in variability.

In ranking methodologies, the lack of a ``true model'' for ranking, makes impossible the calculation of an accuracy indicator \cite{Lin-2017}, however, the patterns analyzed indicates that the WSPAEF is the best option to proceed with the model sorting exercise if the Taylor criteria is used as a guideline.

\begin{figure}[p]
  \centering
  \includegraphics[scale=0.85]{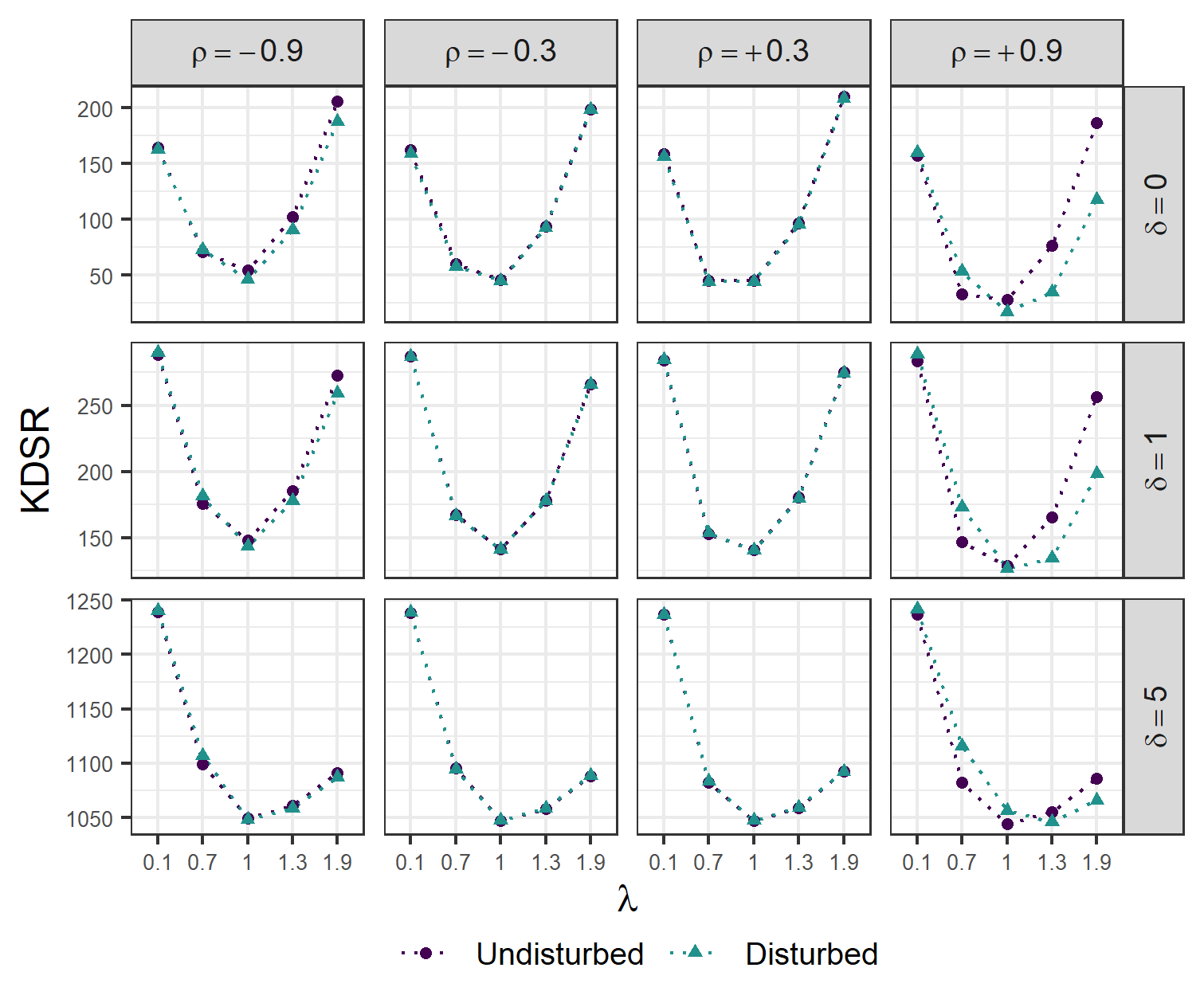}
  \caption{Kolmogorov distance}
    \caption*{Notes: $\rho$: linear correlation, $\lambda$: variation ratio, $\delta$: bias. The ``Undisturbed'' series refers to normal data while the ``Disturbed'' series refers to a log-normal transformation of this data. Lower values indicates a better model.}
  \label{fig:ksd}
\end{figure}

\subsection{Application}

The chosen indicator will be utilized for evaluating the monthly outputs of precipitation and temperature variables from 48 CMIP6 models. For precipitation, the reference data is part of the Global Precipitation Climatology Project version 2.3 dataset \cite{Adler-2003}. These data include measurements from sensors and satellites from 1979 to the present, covering a global grid (88.75\degree N-88.75\degree S, 1.25\degree E-358.75\degree E) with a resolution of 2.5\degree. The temperature data is obtained from reanalysis data provided by the National Centers for Environmental Prediction (NCEP) and the National Center for Atmospheric Research (NCAR) \cite{Kalnay-1996}. The temporal coverage is from 1948 to the present, and the spatial coverage is 90\degree N-90\degree S, 0\degree E-357.5\degree E, with a resolution of 2.5\degree. A sea surface temperature dataset from the National Oceanic and Atmospheric Administration (NOAA) \cite{Huang-2017} is also used to gauge El Niño South Oscilation (ENSO) and Tropical North Atlantic (TNA) teleconnection patterns. It contains temperature measurements from ships and buoys from 1854 to the present, covering the area from 88\degree N-88\degree S, 0\degree E-358\degree E, with a resolution of 2.0\degree. The model data was obtained from the World Climate Research Program (WCRP). The complete list of models can be seen in Table \ref{tbl:models}. The model run indicates the realization (r), initialization method (i), physics (p), and forcing (f) used. All the models were transformed to a 2.5\degree resolution to match the observed data using bilinear interpolation.

\begin{figure}[ht]
  \centering
  \includegraphics{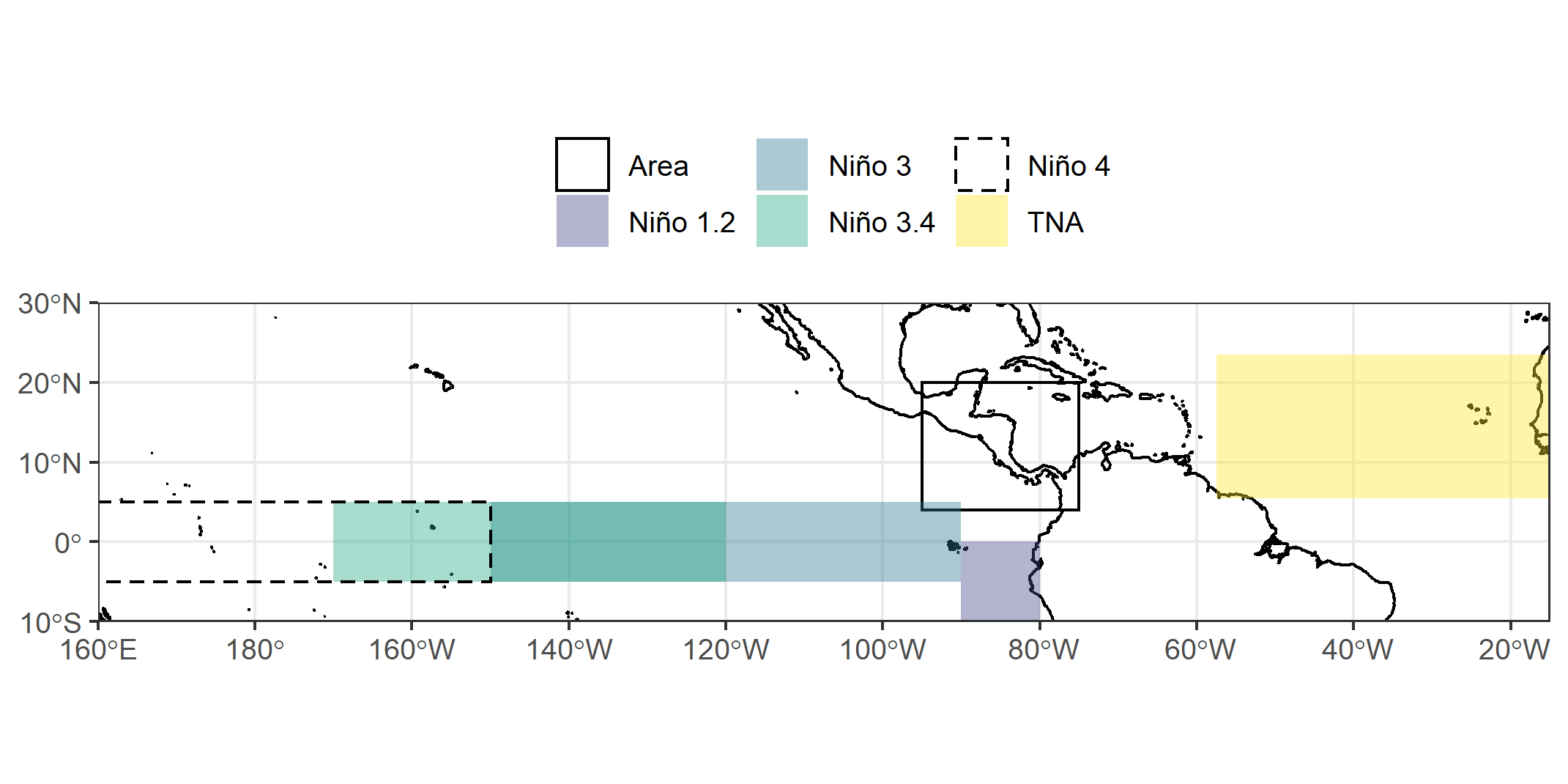}
  \caption{Geographical area under study}
  \label{fig:map}
\end{figure}

The area under study (4\degree N-20\degree N, 95\degree W-75\degree W), as well as the teleconnection regions can be seen in Figure \ref{fig:map}. The temporal frame of the study is limited to the period from 1979 to 1999. 

The comparison will focus on the annual seasonal cycles rather than the entire spatio-temporal field, resulting in the derivation of new variables. To illustrate, by computing the mean precipitation values for each grid point in January and repeating this process for the other months, we can obtain the annual cycle of mean precipitation. In this study, in addition to the provided example, the annual cycle of standard deviations for precipitation and their analogues for surface temperature were also used. The resulting derived variables are: precipitation mean (PRM), precipitation standard deviation (PRS), temperature mean (TPM) and temperature standard deviation (TPS). The teleconnection patterns were generated using the procedure proposed by Hidalgo and Alfaro (2015). For each season (December-January-February [DJF], March-April-May [MAM], June-July-August [JJA], September-October-November [SON]), the temporal correlation between the sea surface temperature time series and the precipitation series within the study region is calculated. These correlations are then combined to form teleconnection variables such as ENSO (El Niño-Southern Oscillation) and TNA (Tropical North Atlantic). The WSPAEF values for each month or season in the annual cycle can be combined using the euclidean norm.

\begin{table}[htbp]
\caption{Preselected model and runs}
\label{tbl:models}
\resizebox{\textwidth}{!}{%
\begin{tabular}{clclc}
\toprule
No & Model   & Run & Institution                                                             & Resolution \\ \midrule
1  & CMCC-CM2-HR4    & r1i1p1f1 & \begin{tabular}[c]{@{}l@{}}Fondazione Centro Euro-Mediterraneo sui \\ Cambiamenti Climatici (FCEMCC)\end{tabular}    & 111 km     \\
2  & CMCC-CM2-VHR4   & r1i1p1f1 & FCEMCC                                                                    & 25 km      \\
3  & CNRM-CM6-1      & r1i1p1f2 & CNRM-CERFACS                                                              & 250 km     \\
4  & CNRM-CM6-1      & r2i1p1f2 & CNRM-CERFACS                                                              & 250 km     \\
5 &
  CNRM-CM6-1 &
  r3i1p1f2 &
  \begin{tabular}[c]{@{}l@{}}Centre National de Recherches Meteorologiques (CNRM),  \\ Centre Europeen de Recherche et de Formation Avancee \\ en Calcul Scientifique (CERFACS)\end{tabular} &
  250 km \\
6  & CNRM-CM6-1-HR   & r1i1p1f2 & CNRM-CERFACS                                                              & 50 km      \\
7  & EC-Earth3P      & r1i1p2f1 & EC-Earth consortium                                                       & 100 km     \\
8  & EC-Earth3P      & r2i1p2f1 & EC-Earth                                                                  & 100 km     \\
9  & EC-Earth3P      & r3i1p2f1 & EC-Earth                                                                  & 50 km      \\
10 & EC-Earth3P-HR   & r1i1p2f1 & EC-Earth                                                                  & 50 km      \\
11 & EC-Earth3P-HR   & r2i1p2f1 & EC-Earth                                                                  & 50 km      \\
12 & EC-Earth3P-HR   & r3i1p2f1 & EC-Earth                                                                  & 50 km      \\
13 & ECMWF-IFS-HR    & r1i1p1f1 & \begin{tabular}[c]{@{}l@{}}European Centre for Medium-Range \\ Weather Forecasts (ECMWF)\end{tabular}                & 25 km      \\
14 & ECMWF-IFS-HR    & r2i1p1f1 & ECMWF                                                                     & 25 km      \\
15 & ECMWF-IFS-HR    & r3i1p1f1 & ECMWF                                                                     & 25 km      \\
16 & ECMWF-IFS-HR    & r4i1p1f1 & ECMWF                                                                     & 25 km      \\
17 & ECMWF-IFS-HR    & r5i1p1f1 & ECMWF                                                                     & 25 km      \\
18 & ECMWF-IFS-HR    & r6i1p1f1 & ECMWF                                                                     & 25 km      \\
19 & ECMWF-IFS-LR    & r1i1p1f1 & ECMWF                                                                     & 50 km      \\
20 & ECMWF-IFS-LR    & r2i1p1f1 & ECMWF                                                                     & 50 km      \\
21 & ECMWF-IFS-LR    & r3i1p1f1 & ECMWF                                                                     & 50 km      \\
22 & ECMWF-IFS-LR    & r4i1p1f1 & ECMWF                                                                     & 50 km      \\
23 & ECMWF-IFS-LR    & r5i1p1f1 & ECMWF                                                                     & 50 km      \\
24 & ECMWF-IFS-LR    & r6i1p1f1 & ECMWF                                                                     & 50 km      \\
25 & ECMWF-IFS-LR    & r7i1p1f1 & ECMWF                                                                     & 50 km      \\
26 & ECMWF-IFS-LR    & r8i1p1f1 & ECMWF                                                                     & 50 km      \\
27 & ECMWF-IFS-MR    & r1i1p1f1 & ECMWF                                                                     & 50 km      \\
28 & ECMWF-IFS-MR    & r2i1p1f1 & ECMWF                                                                     & 50 km      \\
29 & ECMWF-IFS-MR    & r3i1p1f1 & ECMWF                                                                     & 50 km      \\
30 &
  GFDL-CM4C192 &
  r1i1p1f1 &
  \begin{tabular}[c]{@{}l@{}}Geophysical Fluid   Dynamics Laboratory (NOAA-GFDL)\end{tabular} &
  100 km \\
31 & HadGEM3-GC31-HH & r1i1p1f1 & Natural Environment Research Council (NERC)                               & 50 km      \\
32 & HadGEM3-GC31-HM & r1i1p1f1 & Met Office Hadley Centre (MOHC)                                           & 50 km      \\
33 & HadGEM3-GC31-HM & r1i2p1f1 & NERC                                                                      & 50 km      \\
34 & HadGEM3-GC31-HM & r1i3p1f1 & MOHC                                                                      & 50 km      \\
35 & HadGEM3-GC31-LL & r1i1p1f1 & MOHC                                                                      & 250 km     \\
36 & HadGEM3-GC31-LL & r1i2p1f1 & MOHC                                                                      & 250 km     \\
37 & HadGEM3-GC31-LL & r1i3p1f1 & MOHC                                                                      & 250 km     \\
38 & HadGEM3-GC31-LL & r1i4p1f1 & MOHC                                                                      & 250 km     \\
39 & HadGEM3-GC31-LL & r1i5p1f1 & MOHC                                                                      & 250 km     \\
40 & HadGEM3-GC31-LL & r1i6p1f1 & MOHC                                                                      & 250 km     \\
41 & HadGEM3-GC31-LL & r1i7p1f1 & MOHC                                                                      & 250 km     \\
42 & HadGEM3-GC31-LL & r1i8p1f1 & MOHC                                                                      & 250 km     \\
43 & HadGEM3-GC31-MM & r1i1p1f1 & MOHC                                                                      & 100 km     \\
44 & HadGEM3-GC31-MM & r1i2p1f1 & MOHC                                                                      & 100 km     \\
45 & HadGEM3-GC31-MM & r1i3p1f1 & MOHC                                                                      & 100 km     \\
46 & INM-CM5-H       & r1i1p1f1 & \begin{tabular}[c]{@{}l@{}}Institute for Numerical Mathematics (INM-RAS)\end{tabular} & 100 km     \\
47 & MPI-ESM1-2-HR   & r1i1p1f1 & Max Planck Institute for Meteorology (MPIM)                               & 100 km     \\
48 & MPI-ESM1-2-XR   & r1i1p1f1 & MPIM                                                                      & 50 km \\ \bottomrule    
\end{tabular}
}
\end{table}

\subsection{Model Ranking}

Given that the performance of each model is assessed based on multiple derived variables, the selection and ordering of these variables present a multicriteria decision problem. In the realm of Global Climate Model (GCM) evaluation, there is no consensus on a specific method. Therefore, three different methods were employed: the Euclidean norm (L2N) as a control, the Technique for Order Preference by Similarity to Ideal Solution (TOPSIS) proposed by \cite{Lai-1994}, and the Preference Ranking Organization Method for Enrichment of Evaluations (PROMETHEE) introduced by \cite{Brans-1986}. These choices were made based on their popularity, properties, different approaches \cite{Salabun-2020}, and previous use in similar studies \cite{Raju-2014, Sithara-2022, Thakur-2022, Zamani-2019}. Finally, the coincident models within the top eight positions according to each of these final rankings were selected.

The following describes each of these methods. Let $A = \{a_{1}, a_{2},...,a_{n}\}$ be a set of $n$ models or alternatives, and $F = \{f_{1},f_{2},...,f_{m}\}$ be a set of $m$ variables or criteria that should be minimized. They are organized as follows:

$$
 \begin{pmatrix}
  f_{1}(a_{1}) & f_{2}(a_{1}) & \cdots & f_{m}(a_{1}) \\
  f_{1}(a_{2}) & f_{2}(a_{2}) & \cdots & f_{m}(a_{2}) \\
  \vdots  & \vdots  & \ddots & \vdots  \\
  f_{1}(a_{n}) & f_{2}(a_{n}) & \cdots & f_{m}(a_{n}) 
 \end{pmatrix}
$$

\subsubsection{Euclidean Norm (L2N)}

The concept of Euclidean norm or L2-norm is used to calculate the distance from a model $a$ to the origin or point 0 in an $m$-dimensional space, as follows:

$$
d_{a} = \sqrt{\sum_{j = 1}^{m} (f_{j}(a))^2}
$$

These distance values are used for ranking operation; the best models are those located at a shorter distance from the origin in the $\mathbbm{R}^m$ space.

\subsubsection{TOPSIS}

This method also relies on Euclidean distance, but instead of using 0 as a reference point, it uses the positive and negative ideal solutions for each variable, represented by $f_{j}(\cdot)^+$ and $f_{j}(\cdot)^-$, respectively, which correspond to the minimum and maximum values of each column. The distances of each model to these values are then calculated as:

$$d_{a}^+ = \sqrt{\sum_{j = 1}^{m} (f_{j}(a)-f_{j}(\cdot)^+)^2}$$

$$d_{a}^- = \sqrt{\sum_{j = 1}^{m} (f_{j}(a)-f_{j}(\cdot)^-)^2}$$

 Then, the normalized score $\xi_{a} = d_{a}^-/(d_{a}^- + d_{a}^+)$ is used to rank the alternatives. A perfect value of 1 will be assigned to the best alternative and 0 to the worst one, so the models with higher scores are considered of superior performance and occupy a lower position in the ranking. The complete description of this process can be found in Lai et al. \cite{Lai-1994}.

\subsubsection{PROMETHEE}

This method compares two alternatives $\{a,a^*\} \in A$ using a preference function $P_{j}(a,a^*)$. This function depends on the difference between the evaluations of $a$ and $a^*$, expressed as $\Delta_{j}(a,a^*) = f_{j}(a) - f_{j}(a^*)$. Although there are various types of preference functions, this study opted for the ``usual function'' \cite{Brans-1986}, described as:

$$
P_{j}(a,a^*)=
  \begin{cases}
    0  & \quad \text{if } \Delta_{j}(a,a^*) \geq 0\\
    1  & \quad \text{if } \Delta_{j}(a,a^*) < 0 \\
  \end{cases}
$$

When using this function, it indicates that $a$ is strictly preferred to $a^*$ for any negative $\Delta_{j}(a,a^*)$. Model ranking is performed based on the score $\Phi(a)$, defined as:

$$
\Phi(a) = \frac{\sum_{a^* \in A} \pi(a,a^*)}{n-1} - \frac{\sum_{a^* \in A} \pi(a^*,a)}{n-1}
$$

where $\pi(a,a^*)$ represents the weighted average of the preference functions:

$$
\pi(a,a^*) = \frac{\sum_{j = 1}^{m} \omega_{j}P_{j}(a,a^*)}{\sum_{j = 1}^{m} \omega_{j}}
$$

Similar to TOPSIS, higher score values are associated with better models. For a detailed version of the general procedure, refer to Brans et al. \cite*{Brans-1986}.

\section{Results}

Table \ref{tab:rankings} presents the ranking for each of the six derived variables, along with the three final rankings obtained using the L2N, TOPSIS, and PROMETHEE methods. The table is not sorted based on the result of a particular method since none can be considered ``true''. When evaluating the Spearman correlations of TOPSIS and PROMETHEE with respect to L2N, correlation values of 0.96 and 0.89 were obtained, respectively, while the correlation between TOPSIS and PROMETHEE was 0.86. These values indicate that the results are similar and consistent. If this were not the case, each ranking method could be questioned \cite{Salabun-2020}.

\begin{table}[htbp]
\caption{Model position by variable and ranking method}
\label{tab:rankings}
\resizebox{\textwidth}{!}{%
\begin{tabular}{cllccccccccc}
\toprule
No. & Model & Run & PRM & PRS & TPM & TPS & ENSO & TNA & L2N & TOPSIS & PROMETHEE \\ \hline
1  & CMCC-CM2-HR4    & r1i1p1f1 & 48 & 47 & 35 & 26 & 5  & 4  & 47 & 48 & 32 \\
2  & CMCC-CM2-VHR4   & r1i1p1f1 & 47 & 46 & 36 & 4  & 47 & 48 & 48 & 47 & 48 \\
3  & CNRM-CM6-1      & r1i1p1f2 & 25 & 7  & 41 & 27 & 9  & 7  & 25 & 28 & 12 \\
4  & CNRM-CM6-1      & r2i1p1f2 & 24 & 15 & 38 & 9  & 17 & 27 & 30 & 30 & 16 \\
5  & CNRM-CM6-1      & r3i1p1f2 & 26 & 17 & 40 & 45 & 28 & 22 & 34 & 35 & 38 \\
6  & CNRM-CM6-1-HR   & r1i1p1f2 & 13 & 37 & 48 & 35 & 23 & 16 & 36 & 36 & 35 \\
7  & EC-Earth3P      & r1i1p2f1 & 29 & 43 & 31 & 36 & 13 & 21 & 33 & 34 & 36 \\
8  & EC-Earth3P      & r2i1p2f1 & 31 & 45 & 34 & 43 & 39 & 24 & 40 & 42 & 44 \\
9  & EC-Earth3P      & r3i1p2f1 & 28 & 42 & 29 & 44 & 2  & 17 & 24 & 32 & 30 \\
10 & EC-Earth3P-HR   & r1i1p2f1 & 2  & 40 & 28 & 2  & 27 & 37 & 14 & 18 & 20 \\
11 & EC-Earth3P-HR   & r2i1p2f1 & 4  & 41 & 1  & 16 & 1  & 13 & 1  & 1  & 2  \\
12 & EC-Earth3P-HR   & r3i1p2f1 & 3  & 39 & 27 & 11 & 14 & 9  & 9  & 15 & 10 \\
13 & ECMWF-IFS-HR    & r1i1p1f1 & 1  & 23 & 18 & 17 & 4  & 14 & 2  & 4  & 3  \\
14 & ECMWF-IFS-HR    & r2i1p1f1 & 19 & 21 & 24 & 1  & 20 & 6  & 8  & 9  & 7  \\
15 & ECMWF-IFS-HR    & r3i1p1f1 & 17 & 33 & 23 & 10 & 3  & 1  & 4  & 8  & 6  \\
16 & ECMWF-IFS-HR    & r4i1p1f1 & 18 & 36 & 20 & 14 & 16 & 44 & 16 & 20 & 26 \\
17 & ECMWF-IFS-HR    & r5i1p1f1 & 11 & 16 & 19 & 3  & 25 & 8  & 7  & 5  & 5  \\
18 & ECMWF-IFS-HR    & r6i1p1f1 & 12 & 35 & 22 & 23 & 44 & 39 & 27 & 22 & 37 \\
19 & ECMWF-IFS-LR    & r1i1p1f1 & 36 & 27 & 45 & 31 & 30 & 20 & 39 & 39 & 39 \\
20 & ECMWF-IFS-LR    & r2i1p1f1 & 38 & 29 & 39 & 38 & 31 & 18 & 41 & 40 & 40 \\
21 & ECMWF-IFS-LR    & r3i1p1f1 & 34 & 19 & 37 & 40 & 24 & 46 & 37 & 38 & 41 \\
22 & ECMWF-IFS-LR    & r4i1p1f1 & 44 & 26 & 47 & 19 & 40 & 43 & 46 & 44 & 46 \\
23 & ECMWF-IFS-LR    & r5i1p1f1 & 35 & 30 & 42 & 37 & 42 & 36 & 44 & 41 & 47 \\
24 & ECMWF-IFS-LR    & r6i1p1f1 & 45 & 25 & 46 & 41 & 10 & 47 & 43 & 46 & 43 \\
25 & ECMWF-IFS-LR    & r7i1p1f1 & 42 & 31 & 44 & 32 & 38 & 30 & 45 & 43 & 45 \\
26 & ECMWF-IFS-LR    & r8i1p1f1 & 37 & 34 & 43 & 29 & 21 & 2  & 38 & 37 & 33 \\
27 & ECMWF-IFS-MR    & r1i1p1f1 & 21 & 5  & 21 & 42 & 26 & 23 & 21 & 21 & 22 \\
28 & ECMWF-IFS-MR    & r2i1p1f1 & 33 & 22 & 26 & 6  & 48 & 34 & 35 & 29 & 34 \\
29 & ECMWF-IFS-MR    & r3i1p1f1 & 32 & 18 & 25 & 8  & 8  & 11 & 12 & 19 & 9  \\
30 & GFDL-CM4C192    & r1i1p1f1 & 20 & 28 & 33 & 28 & 7  & 19 & 15 & 27 & 18 \\
31 & HadGEM3-GC31-HH & r1i1p1f1 & 41 & 3  & 15 & 7  & 41 & 28 & 31 & 23 & 19 \\
32 & HadGEM3-GC31-HM & r1i1p1f1 & 40 & 6  & 3  & 5  & 19 & 5  & 18 & 14 & 4  \\
33 & HadGEM3-GC31-HM & r1i2p1f1 & 43 & 9  & 4  & 18 & 22 & 25 & 28 & 26 & 14 \\
34 & HadGEM3-GC31-HM & r1i3p1f1 & 39 & 1  & 6  & 34 & 11 & 41 & 23 & 25 & 17 \\
35 & HadGEM3-GC31-LL & r1i1p1f1 & 9  & 13 & 12 & 15 & 32 & 15 & 5  & 3  & 8  \\
36 & HadGEM3-GC31-LL & r1i2p1f1 & 6  & 32 & 13 & 13 & 37 & 38 & 11 & 11 & 23 \\
37 & HadGEM3-GC31-LL & r1i3p1f1 & 10 & 20 & 17 & 22 & 34 & 33 & 10 & 10 & 21 \\
38 & HadGEM3-GC31-LL & r1i4p1f1 & 16 & 11 & 8  & 25 & 6  & 3  & 3  & 2  & 1  \\
39 & HadGEM3-GC31-LL & r1i5p1f1 & 5  & 38 & 16 & 20 & 18 & 29 & 6  & 7  & 15 \\
40 & HadGEM3-GC31-LL & r1i6p1f1 & 14 & 14 & 9  & 24 & 45 & 35 & 20 & 13 & 24 \\
41 & HadGEM3-GC31-LL & r1i7p1f1 & 8  & 10 & 5  & 12 & 46 & 32 & 13 & 6  & 11 \\
42 & HadGEM3-GC31-LL & r1i8p1f1 & 15 & 44 & 7  & 21 & 36 & 26 & 19 & 16 & 27 \\
43 & HadGEM3-GC31-MM & r1i1p1f1 & 23 & 2  & 10 & 46 & 29 & 45 & 22 & 24 & 28 \\
44 & HadGEM3-GC31-MM & r1i2p1f1 & 27 & 4  & 14 & 30 & 43 & 40 & 26 & 17 & 29 \\
45 & HadGEM3-GC31-MM & r1i3p1f1 & 22 & 8  & 11 & 33 & 35 & 12 & 17 & 12 & 13 \\
46 & INM-CM5-H       & r1i1p1f1 & 30 & 48 & 2  & 39 & 15 & 10 & 29 & 33 & 25 \\
47 & MPI-ESM1-2-HR   & r1i1p1f1 & 7  & 12 & 32 & 48 & 33 & 31 & 32 & 31 & 31 \\
48 & MPI-ESM1-2-XR   & r1i1p1f1 & 46 & 24 & 30 & 47 & 12 & 42 & 42 & 45 & 42\\ \bottomrule
\end{tabular}
}
\end{table}

\begin{figure}[htbp]
  \centering
  \includegraphics{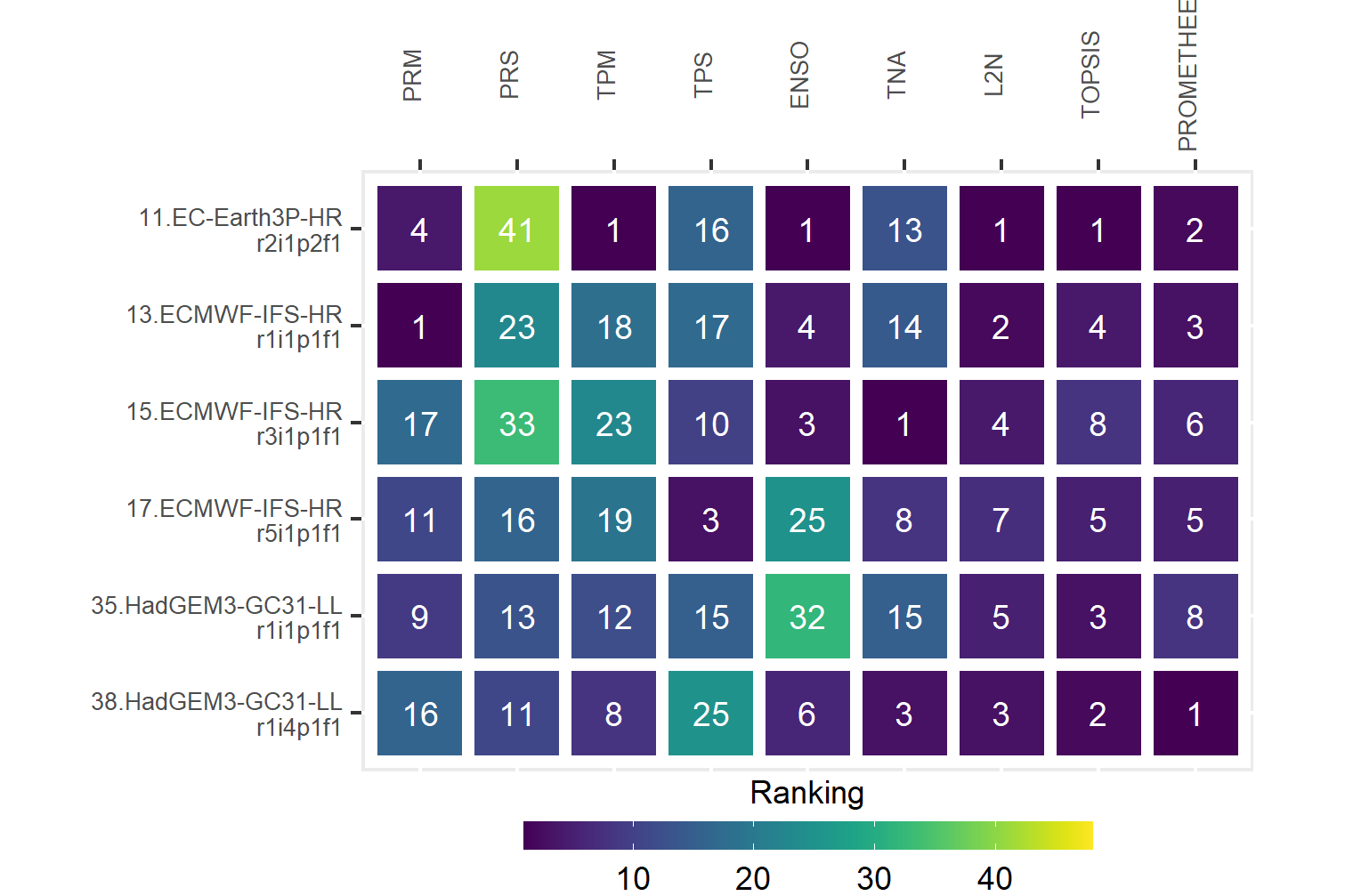}
  \caption{Selected models positions}
  \label{fig:rank6}
\end{figure}

Six models were found to be in common among the top eight positions in all three multi-criteria methods. These are shown in Figure \ref{fig:rank6} with their respective positions. Based on the criteria used, it can be asserted that any of these models are capable of adequately reproducing the annual cycle of the analyzed variables.

Model 11 (EC-Earth3P-HR r2i1p2f1) is the best according to L2N and TOPSIS, and the second best according to PROMETHEE, thanks to its ability to reproduce TPM and ENSO. However, it should be noted that it ranks 41st for the PRS variable. This should be taken into consideration if this variable is of special interest. Other more balanced alternatives are Model 13 (ECMWF-IFS-HR r1i1p1f1) and Model 38 (HadGEM3-GC31-LL r1i4p1f1). Notice that all selected models occupy relatively high positions in at least one variable.

\begin{figure}
    \centering
     \begin{subfigure}[t]{0.48\textwidth}
         \centering
         \includegraphics[width=\textwidth]{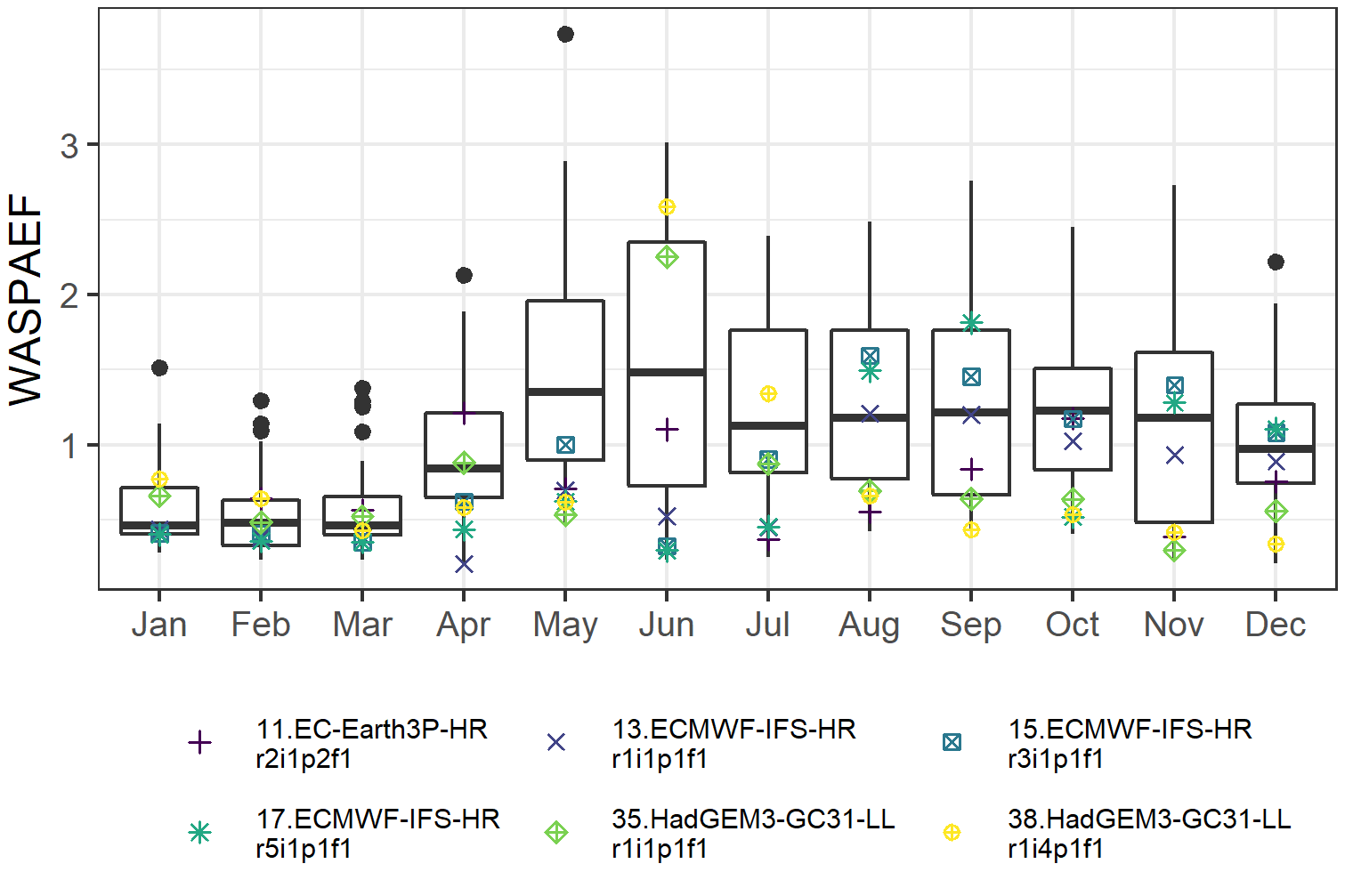}
         \caption{PRM}
         \label{fig:prmeanbx}
     \end{subfigure}
     \begin{subfigure}[t]{0.48\textwidth}
         \centering
         \includegraphics[width=\textwidth]{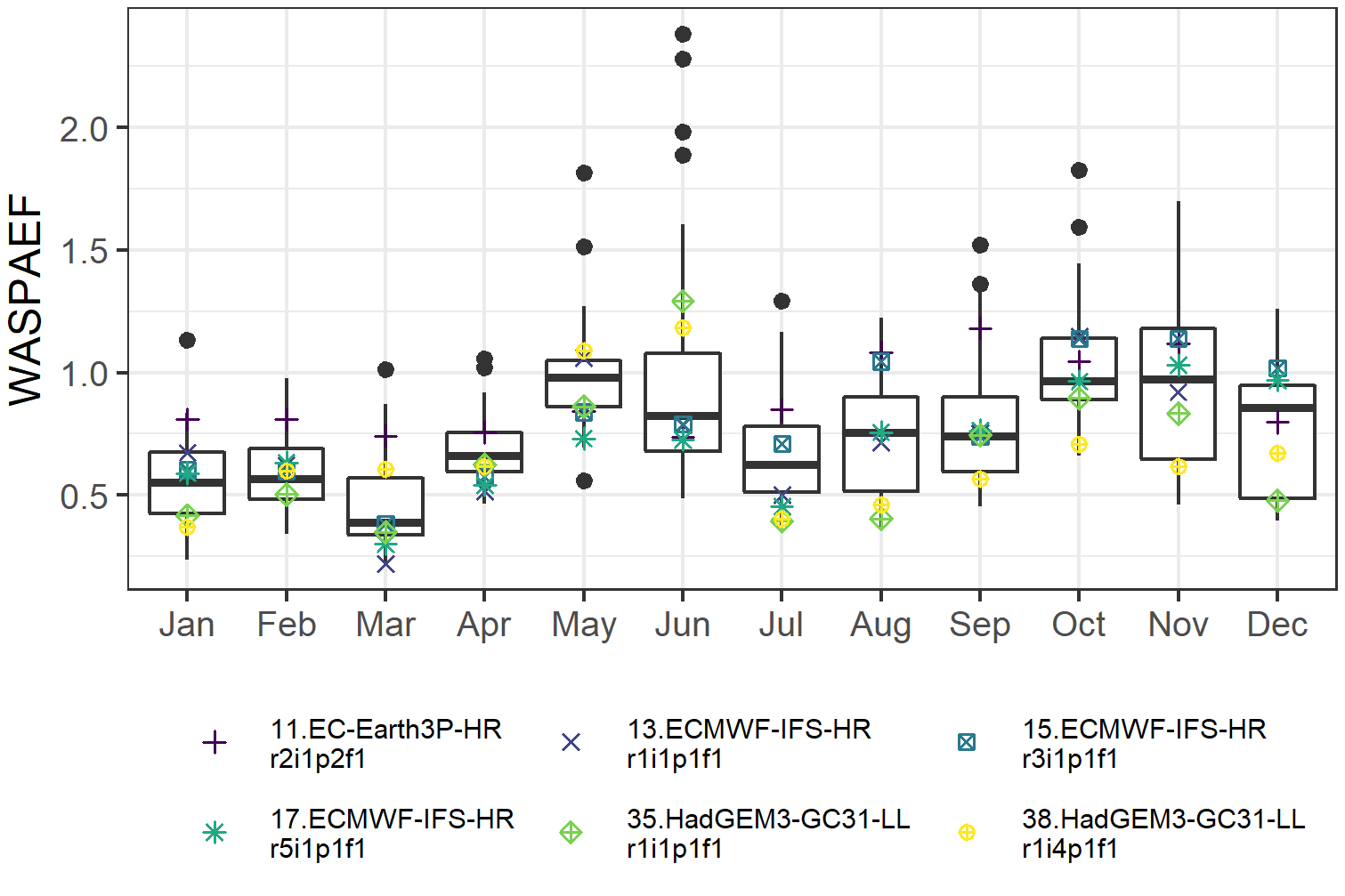}
         \caption{PRS}
         \label{fig:prsdbx}
     \end{subfigure} 
        \par\bigskip
     \begin{subfigure}[t]{0.48\textwidth}
         \centering
         \includegraphics[width=\textwidth]{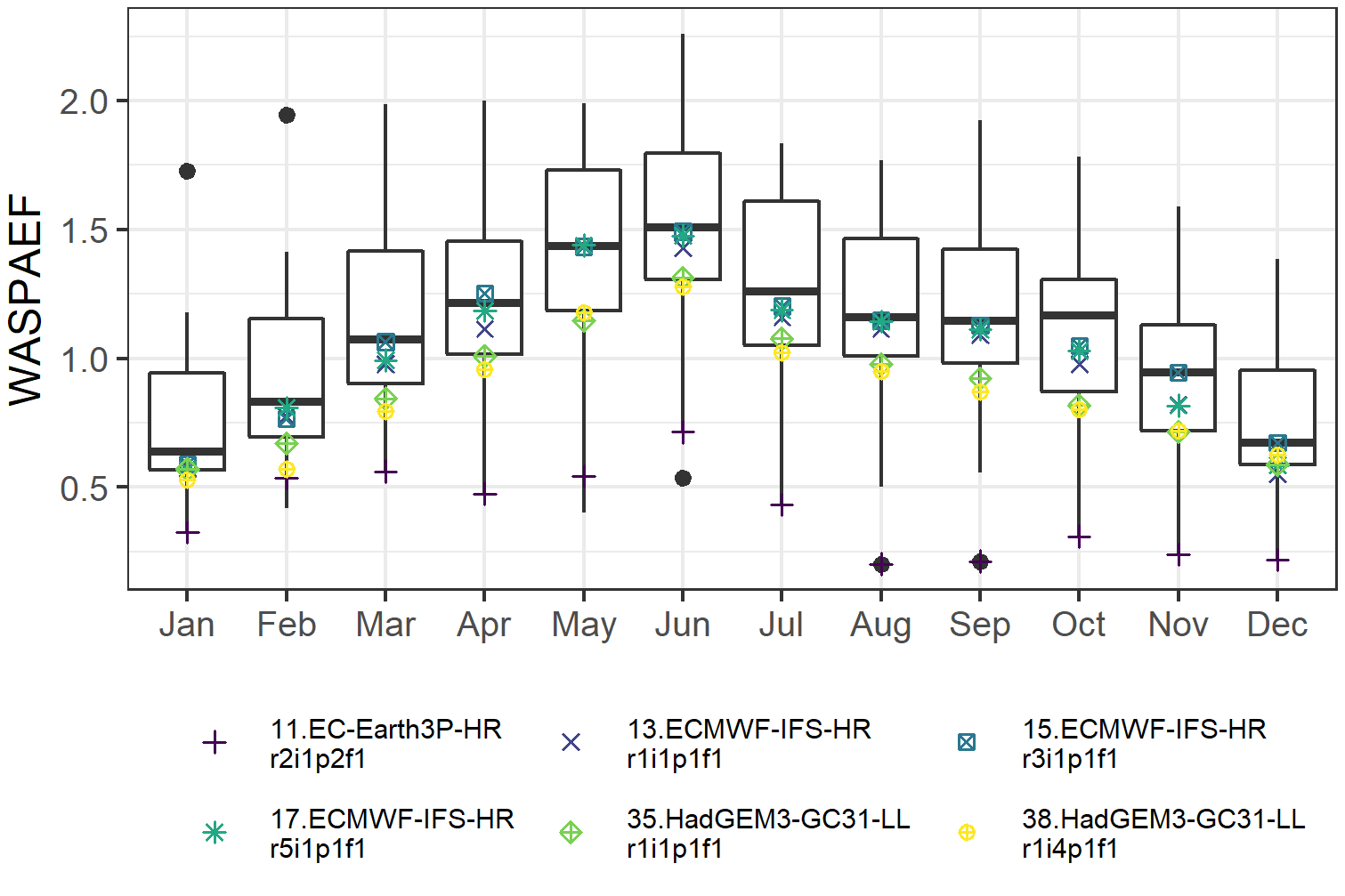}
         \caption{TPM}
         \label{fig:tasmeanbx}
     \end{subfigure}
     \begin{subfigure}[t]{0.48\textwidth}
         \centering
         \includegraphics[width=\textwidth]{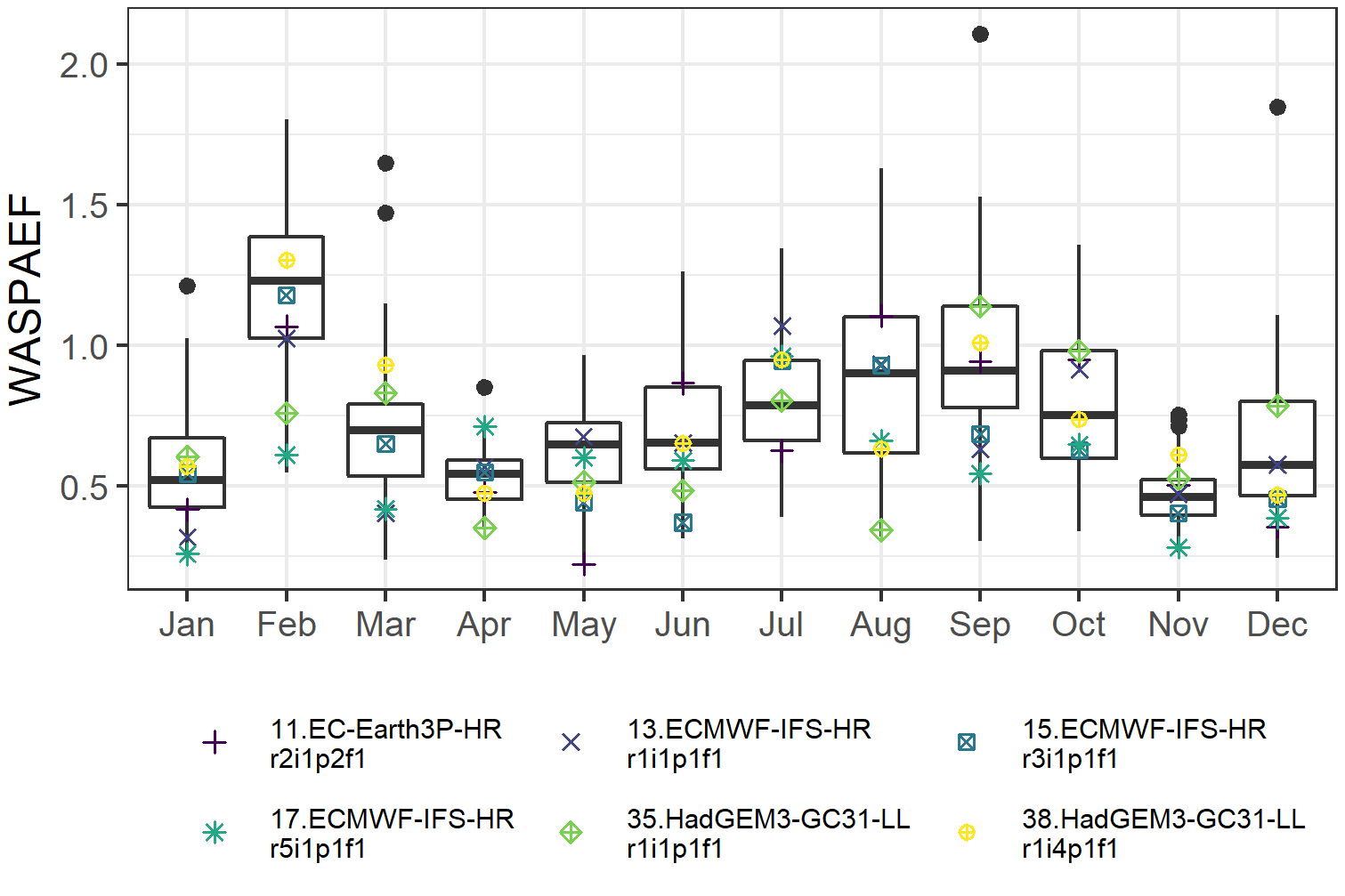}
         \caption{TPS}
         \label{fig:tassdbx}
     \end{subfigure}
    \caption{Precipitation and temperature derived variables WSAPEF by month}
    \label{fig:bxpprtas}
\end{figure}

\begin{figure}
     \centering
     \begin{subfigure}[t]{0.48\textwidth}
         \centering
         \includegraphics[width=\textwidth]{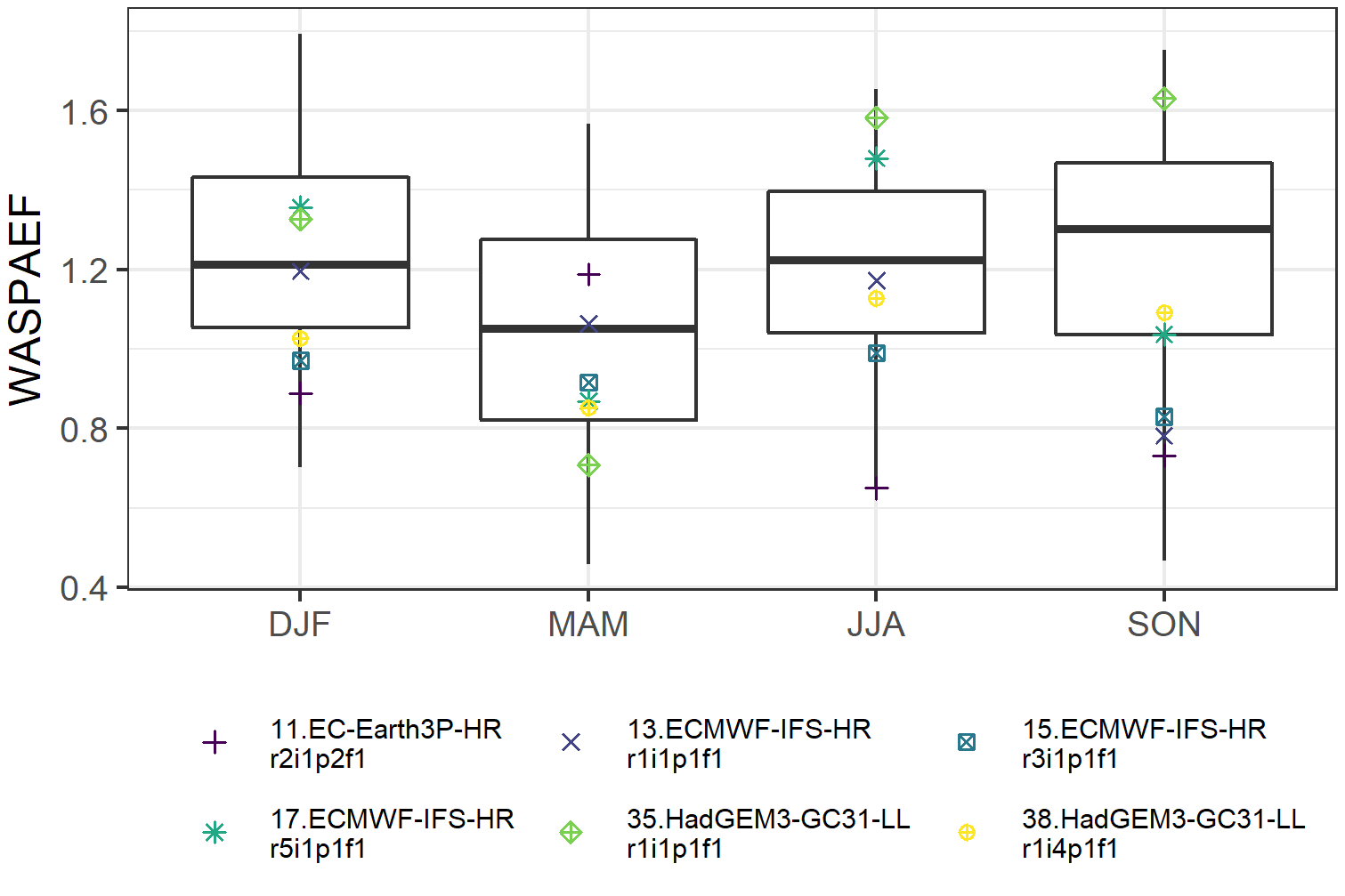}
         \caption{Niño 1.2}
         \label{fig:ensobx}
     \end{subfigure}
     \begin{subfigure}[t]{0.48\textwidth}
         \centering
         \includegraphics[width=\textwidth]{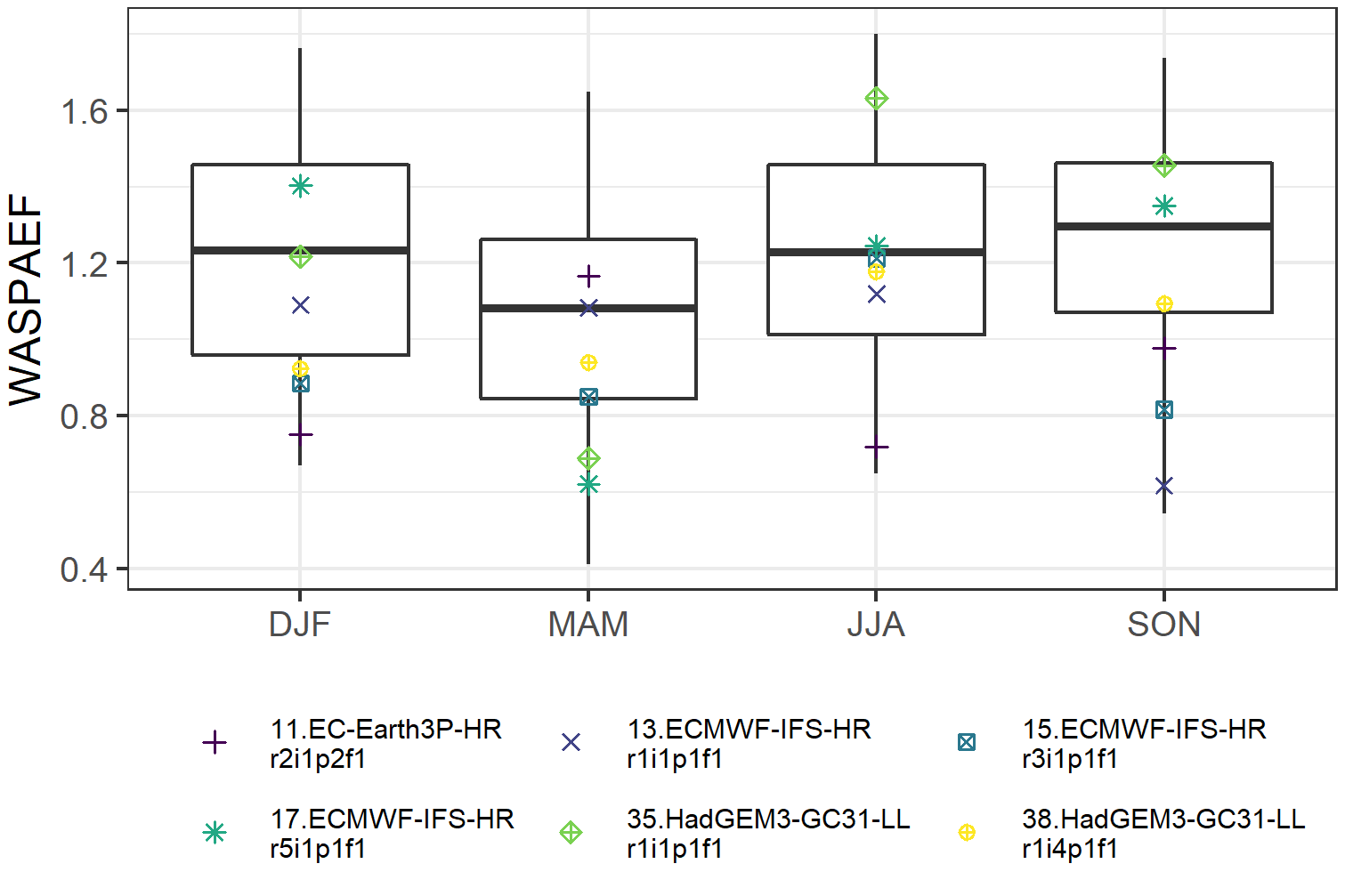}
         \caption{Niño 3}
         \label{fig:enso3dbx}
     \end{subfigure}
    \par\bigskip
    \begin{subfigure}[t]{0.48\textwidth}
         \centering
         \includegraphics[width=\textwidth]{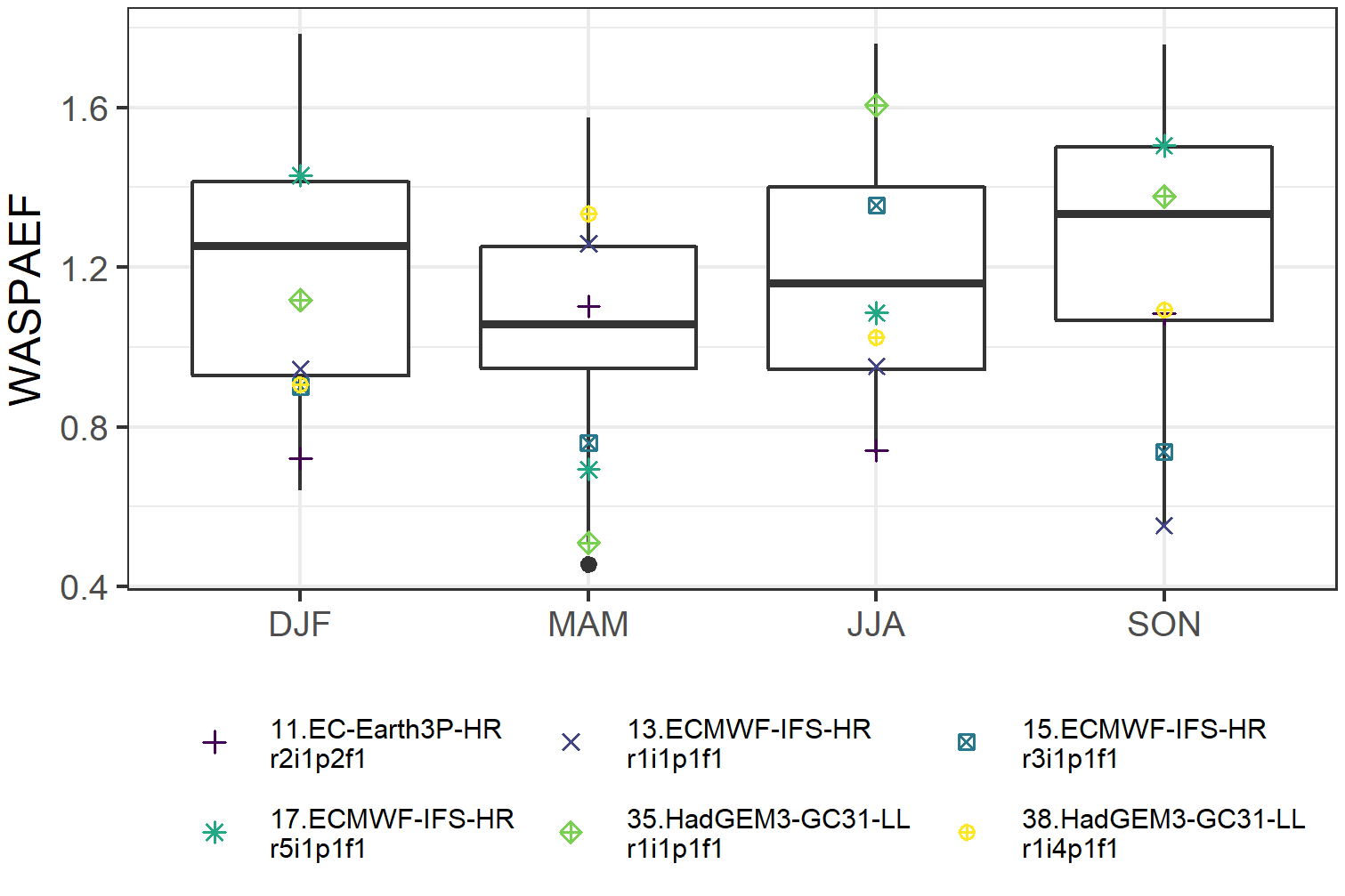}
         \caption{Niño 3.4}
         \label{fig:enso34dbx}
     \end{subfigure}
     \begin{subfigure}[t]{0.48\textwidth}
         \centering
         \includegraphics[width=\textwidth]{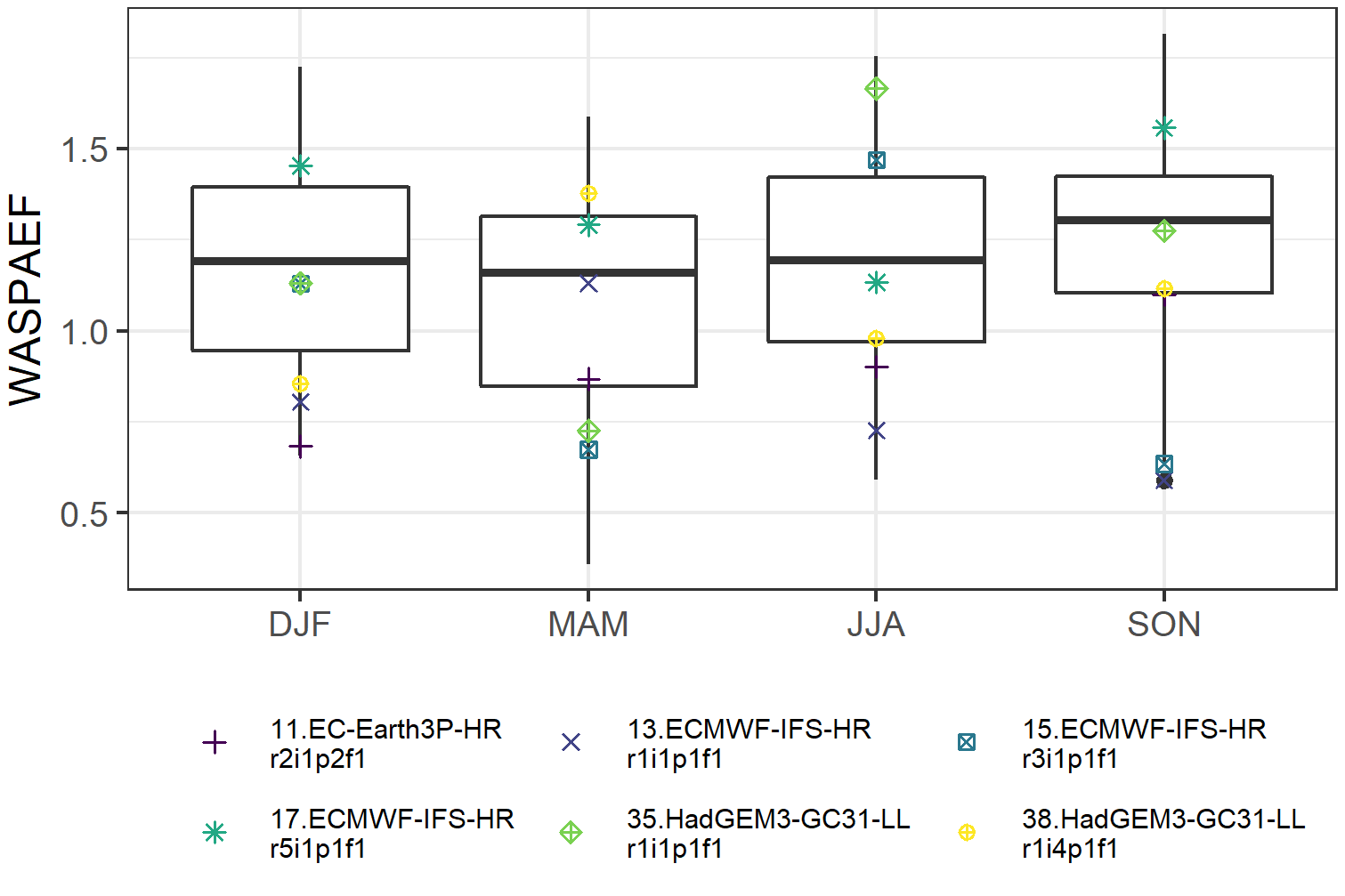}
         \caption{Niño 4}
         \label{fig:enso4dbx}
     \end{subfigure}
     \par\bigskip
    \begin{subfigure}[t]{0.48\textwidth}
         \centering
         \includegraphics[width=\textwidth]{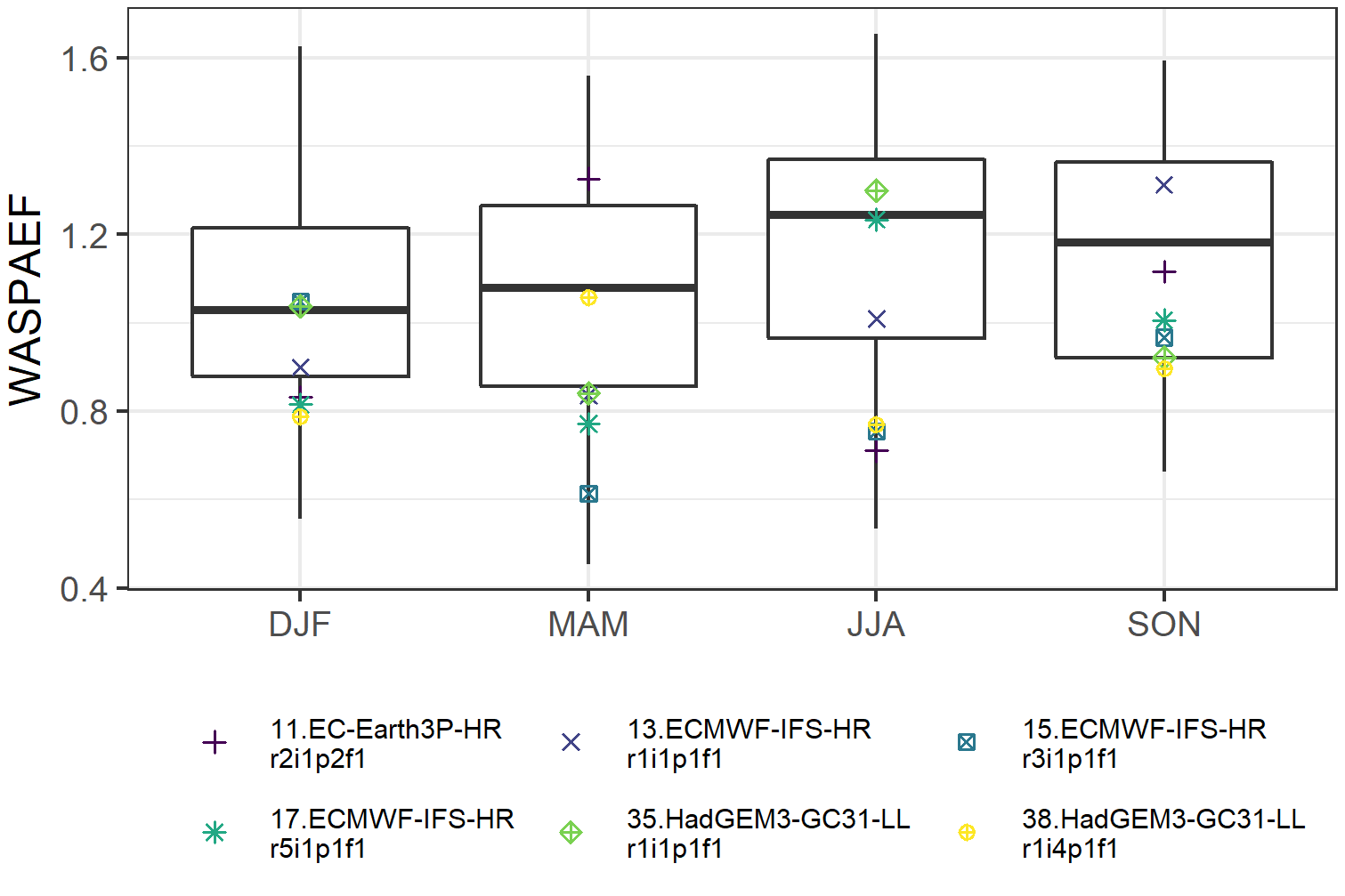}
         \caption{TNA}
         \label{fig:atndbx}
     \end{subfigure}
        \caption{Teleconnection WSPAEF by season}
        \label{fig:bxptlc}
\end{figure}

Figures \ref{fig:bxpprtas} and \ref{fig:bxptlc} show the model performance for the temporal grouping of the annual cycle. For mean precipitation, in general, all models show better performance during the drier months (Jan - Mar) and also less variability among them \cite{Maldonado-2018}. The same pattern is repeated for the standard deviation, although less prominently. For mean temperature, the WSPAEF values consistently increase until reaching a maximum in June and then decrease. On the other hand, the variability of the results seems to remain constant throughout the cycle. It can also be observed that Model 11 consistently ranks at the minimum or near-minimum positions in all months, justifying its position in the ranking for this variable. TPS does not exhibit any clear trend, but it is worth noting its overall good performance for the month of November and the opposite for February.

The teleconnection patterns (Figure \ref{fig:bxptlc}), segregated by region, do not show clear trends and appear to be fairly stable. However, the selected individual models exhibit an interesting phenomenon. Model 11, which ranks first overall for ENSO, seems to perform poorly for the MAM season, whereas Model 35, which ranks last for this variable among the selected models, performs particularly well for this season.

Although other studies have evaluated CMIP6 GCMs in Central America \cite{Almazroui-2021,Zhang2-2022}, differences in the choice of spatial and temporal domains, variables under analysis, reference datasets, and performance indicators, among others, prevent a direct comparison. Additionally, the preselection of models and their respective runs was not the same, further reducing the likelihood of coincidences. With this in mind, it is important to highlight that at least one model from the EC-Earth consortium (EC-Earth3P-HR) shows superior performance in this study and previous studies (EC-Earth3 and EC-Earth3-Veg).

\section{Conclusions}

Methods for evaluating model performance are necessary in the atmospheric sciences, where multiple models are constructed to reproduce a climatic variable, and it is important to identify the ones that resemble the reference data. The qualifier ``best'' can be ambiguous, so defining a ``good model'' is fundamental, as is determining whether these methods can detect its characteristics.

In this research, it was assumed that a good model is one that generates a field with a high linear correlation with a reference field, as well as similar values for mean, dispersion, and distribution shape. The behavior of six techniques was examined across various scenarios involving these attributes, with three of them being novel in the literature for this particular application. Among these techniques, the WSPAEF, a multi-component measure, demonstrated the highest proficiency in accurately identifying the best alternatives.

A list of 48 CMIP6 models was evaluated with this indicator, based on their ability to reproduce the annual cycle of mean and standard deviation of precipitation and temperature in Central America during the last two decades of the 20th century, as well as the teleconnection patterns in various ENSO and TNA regions. Six models were selected based on three multi-criteria methods. These are: EC-Earth3P-HR (r2i1p2f1), ECMWF-IFS-HR (r1i1p1f1), ECMWF-IFS-HR (r3i1p1f1), ECMWF-IFS-HR (r5i1p1f1), HadGEM3-GC31-LL (r1i1p1f1), HadGEM3-GC31-LL (r1i4p1f1).

When analyzing them individually by variable and month or season, it was identified that their performance is not uniform. For example, the best-performing model overall (EC-Earth3P-HR r2i1p2f1) exhibits poor performance in reproducing the standard deviation of precipitation or the teleconnection patterns during the period from March to May. These performance differences between variables or time periods lead to considering the use of an ensemble of models \cite{SMHI}. The task of reducing the size and precision of these ensembles can be performed using the procedure proposed in this study.

Despite the fact that more intricate methods based on Functional Data Analysis (FDA) did not yield satisfactory outcomes in this study, it is not reasonable to discard this approach. The quest for a metric or procedure enabling a more direct quantification of similarity, devoid of the necessity to aggregate data using statistics like mean and standard deviation, as well as the pursuit of more sophisticated statistical techniques to assess relationships between distinct regions (teleconnections), persist as open challenges.

In addition to introducing the proposed measure and selection method, this work also sheds light on the limitations of other well-established techniques. By doing so, it aims to stimulate the development of new evaluation methods specifically designed for these indicators. 

\section*{Data availability}

The reference  data is available at \url{https://psl.noaa.gov/data/gridded/tables/surface.html} while the models data can be found at \url{https://esgf-node.llnl.gov/search/cmip6/}. 

\section*{Funding}

Not applicable

\section*{Conflict of interest}
The authors declare no conflict of interest.

\section*{Ethics approval}

Not applicable

\section*{Patient consent}

Not applicable

\section*{Permission to reproduce material from other sources}

Not applicable

\section*{Clinical trial registration}

Not applicable

\section*{Supporting Information}

\clearpage
\printbibliography

@article{Panaretos-2019,
author = {Panaretos, Victor M. and Zemel, Yoav},
title = {Statistical Aspects of Wasserstein Distances},
journal = {Annual Review of Statistics and Its Application},
volume = {6},
number = {1},
pages = {405-431},
year = {2019},
doi = {10.1146/annurev-statistics-030718-104938},
URL = { 
        https://doi.org/10.1146/annurev-statistics-030718-104938},
eprint = { 
        https://doi.org/10.1146/annurev-statistics-030718-104938}
}

@article{Vissio-2020,
author = {Vissio, Gabriele and Lembo, Valerio and Lucarini, Valerio and Ghil, Michael},
title = {Evaluating the Performance of Climate Models Based on Wasserstein Distance},
journal = {Geophysical Research Letters},
volume = {47},
number = {21},
pages = {e2020GL089385},
doi = {https://doi.org/10.1029/2020GL089385},
url = {https://agupubs.onlinelibrary.wiley.com/doi/abs/10.1029/2020GL089385},
eprint = {https://agupubs.onlinelibrary.wiley.com/doi/pdf/10.1029/2020GL089385},
note = {e2020GL089385 10.1029/2020GL089385},
year = {2020}
}

@article{Tan-2017,
    author = "Tan, Chengguo and Yan, Suiyu",
    title = "{Spatiotemporal data organization and application research}",
    journal = "The International Archives of the Photogrammetry, Remote Sensing and Spatial Information Sciences",
    volume = "42",
    number = "2",
    pages = "1363--1366",
    year = "2017",
    url = "https://doaj.org/article/b52c28323dec4e118ba78364aade3fe9"
}

@article{Kalnay-1996,
    author = "Kalnay, E. and Kanamitsu, M. and Kistler, R. and Collins, W. and Deaven, D. and Gandin, L. and Iredell, M. and Saha, S. and White, G. and Woollen, J. and Zhu, Y. and Chelliah, M. and Ebisuzaki, W. and Higgins, W. and Janowiak, J. and Mo, K.C. and Ropelewski, C. and Wang, J. and Leetmaa, A. and Reynolds, R. and Jenne, Roy and Joseph, Dennis",
    title = "{The NCEP/NCAR 40-Year Reanalysis Project}",
    journal = "Bulletin of the American Meteorological Society",
    volume = "77",
    number = "3",
    pages = "437--472",
    year = "1996",
    url = "https://doi.org/10.1175/1520-0477(1996)077<0437:TNYRP>2.0.CO;2",
}

@article{Adler-2003,
    author = "Adler, Robert F. and Huffman, George J. and Chang, Alfred and Ferraro, Ralph and Xie, Ping-Ping and Janowiak, John and Rudolf, Bruno and Schneider, Udo and Curtis, Scott and Bolvin, David and Gruber, Arnold and Susskind, Joel and Arkin, Philip and Nelkin, Eric",
    title = "{The Version-2 Global Precipitation Climatology Project (GPCP) Monthly Precipitation Analysis (1979–Present)}",
    journal = "Journal of Hydrometeorology",
    volume = "4",
    number = "6",
    pages = "1147--1167",
    year = "2003",
    url = "https://doi.org/10.1175/1525-7541(2003)004<1147:TVGPCP>2.0.CO;2",
}

@dataset{Huang-2017,
    author = {Huang, Boyin and Thorne, Peter W. and Banzon, Viva F. and Boyer, Tim and Chepurin, Gennady and Lawrimore, Jay H. and Menne, Matthew J. and Smith, Thomas M. and Vose, Russell S. and Zhang, Huai-Min}, 
    title = {NOAA Extended Reconstructed Sea Surface Temperature (ERSST), Version 5},
    publisher = {NOAA National Centers for Environmental Information},
    year = {2017},
    url = {https://psl.noaa.gov/data/gridded/data.noaa.ersst.v5.html}
}

@article{Ramsay-1982,
    author = "Ramsay, James",
    title = "{When the data are functions}",
    journal = "Psychometrika",
    volume = "47",
    pages = "379--396",
    year = "1982",
    url = "https://doi.org/10.1007/BF02293704"
}

@article{Suhaila-2017,
    author = "Suhaila, Jamaludin and Yusop, Zulkifli",
    title = "{Spatial and temporal variabilities of rainfall data using functional data analysis}",
    journal = "Theoretical and Applied Climatology",
    volume = "129",
    number = "1",
    pages = "229--242",
    year = "2017",
    url = "https://doi.org/10.1007/s00704-016-1778-x"
}

@article{Lin-2017,
    author = "Lin, Zhuhua and Zhou, Yingchun",
    title = "{Ranking of functional data in application to worldwide $PM_{10}$ data analysis}",
    journal = "Environmental and Ecological Statistics",
    volume = "24",
    number = "4",
    pages = "469--484",
    url = "https://doi.org/10.1007/s10651-017-0384-0",
    year = "2017"
}

@article{Murphy-1988,
    author = "Murphy, Allan H.",
    title = "{Skill Scores Based on the Mean Square Error and Their Relationships to the Correlation Coefficient}",
    journal = "Monthly Weather Review",
    volume = "116",
    number = "12",
    pages = "2417--2424",
    year = "1988",
    url = "https://doi.org/10.1175/1520-0493(1988)116<2417:SSBOTM>2.0.CO;2",
}

@article{Eyring-2016,
    author = "Eyring, V. and Bony, S. and Meehl, G. A. and Senior, C. A. and Stevens, B. and Stouffer, R. J. and Taylor, K. E",
    title = "{Overview of the Coupled Model Intercomparison Project Phase 6 (CMIP6) experimental design and organization}",
    journal = "Geoscientific Model Development",
    volume = "9",
    pages = "1937--1958",
    year = "2016",
    url = "https://doi.org/10.5194/gmd-9-1937-2016",
}

@article{Hidalgo-2015,
    author = "Hidalgo, Hugo G. and Alfaro, Eric J.",
    title = "{Skill of CMIP5 climate models in reproducing 20th century basic climate features in Central America}",
    journal = "International Journal of Climatology",
    volume = "35",
    number = "12",
    pages = "3397--3421",
    year = "2015",
    url = "https://rmets.onlinelibrary.wiley.com/doi/abs/10.1002/joc.4216"
}

@article{Hidalgo-2012,
    author = "Hidalgo, Hugo G. and Alfaro, Eric J.",
    title = "{Global Model selection for evaluation of climate change projections in the Eastern Tropical Pacific Seascape}",
    journal = "Revista de Biología Tropical",
    volume = "60",
    number = "3",
    pages = "67--81",
    year = "2012",
    url = "http://revistas.ucr.ac.cr/index.php/rbt/issue/archive"
}

@article{Knutti-2010,
    author = "Knutti, Reto",
    title = "{The end of model democracy?}",
    journal = "Climatic Change",
    volume = "102",
    number = "3",
    pages = "395--404",
    year = "2010",
    url = "https://doi.org/10.1007/s10584-010-9800-2"
}

@article{Taylor-2001,
author = {Taylor, Karl E.},
title = "{Summarizing multiple aspects of model performance in a single diagram}",
journal = {Journal of Geophysical Research: Atmospheres},
volume = {106},
number = {7},
pages = {7183--7192},
url = {https://agupubs.onlinelibrary.wiley.com/doi/abs/10.1029/2000JD900719},
year = {2001}
}

@article{Raju-2014,
    author = {Raju, K. S. and Nagesh K.D.},
    title = "{Ranking of global climate models for India using multicriterion analysis}",
    journal = {Climate Research},
    volume = {60},
    number = {2},
    pages = {103--117},
    year = {2014},
    url = {https://www.int-res.com/abstracts/cr/v60/n2/p103-117/},
}

@article{Gleckler-2008,
author = {Gleckler, P. J. and Taylor, K. E. and Doutriaux, C.},
title = {Performance metrics for climate models},
journal = {Journal of Geophysical Research: Atmospheres},
volume = {113},
number = {D6},
url = {https://agupubs.onlinelibrary.wiley.com/doi/abs/10.1029/2007JD008972},
year = {2008}
}

@article{Harris-2020,
author = {Trevor Harris and Bo Li and Nathan J. Steiger and Jason E. Smerdon and Naveen Narisetty and J. Derek Tucker},
title = {Evaluating Proxy Influence in Assimilated Paleoclimate Reconstructions—Testing the Exchangeability of Two Ensembles of Spatial Processes},
journal = {Journal of the American Statistical Association},
URL = {https://doi.org/10.1080/01621459.2020.1799810},
year  = {2020}
}

@inproceedings{Tukey-1975,
  title={Mathematics and the picturing of data},
  author={J. W. Tukey},
  year={1975}
}

@book{Wikle-2019,
    title = {Spatio-Temporal Statistics with R},
    author = {Wikle, Christopher K. and Zammit-Mangion, Andrew and Cressie, Noel},
    isbn = {9781351769723},
    series = {Springer Series in Statistics},
    year = {2019},
    url = {https://doi.org/10.1201/9781351769723},
    publisher = {Chapman and Hall/CRC}
}

@article{Gupta-2009,
title = {Decomposition of the mean squared error and NSE performance criteria: Implications for improving hydrological modelling},
journal = {Journal of Hydrology},
volume = {377},
number = {1},
pages = {80-91},
year = {2009},
issn = {0022-1694},
doi = {https://doi.org/10.1016/j.jhydrol.2009.08.003},
url = {https://www.sciencedirect.com/science/article/pii/S0022169409004843},
author = {Hoshin V. Gupta and Harald Kling and Koray K. Yilmaz and Guillermo F. Martinez},
}

@article{Kling-2012,
title = {Runoff conditions in the upper Danube basin under an ensemble of climate change scenarios},
journal = {Journal of Hydrology},
volume = {424-425},
pages = {264-277},
year = {2012},
issn = {0022-1694},
doi = {https://doi.org/10.1016/j.jhydrol.2012.01.011},
url = {https://www.sciencedirect.com/science/article/pii/S0022169412000431},
author = {Harald Kling and Martin Fuchs and Maria Paulin},
}

@article{Pincus-2008,
author = {Pincus, Robert and Batstone, Crispian P. and Hofmann, Robert J. Patrick and Taylor, Karl E. and Glecker, Peter J.},
title = {Evaluating the present-day simulation of clouds, precipitation, and radiation in climate models},
journal = {Journal of Geophysical Research: Atmospheres},
volume = {113},
number = {D14},
pages = {},
doi = {https://doi.org/10.1029/2007JD009334},
url = {https://agupubs.onlinelibrary.wiley.com/doi/abs/10.1029/2007JD009334},
eprint = {https://agupubs.onlinelibrary.wiley.com/doi/pdf/10.1029/2007JD009334},
year = {2008}
}

@Article{Centella-2020,
AUTHOR = {Centella-Artola, Abel and Bezanilla-Morlot, Arnoldo and Taylor, Michael A. and Herrera, Dimitris A. and Martinez-Castro, Daniel and Gouirand, Isabelle and Sierra-Lorenzo, Maibys and Vichot-Llano, Alejandro and Stephenson, Tannecia and Fonseca, Cecilia and Campbell, Jayaka and Alpizar, Milena},
TITLE = {Evaluation of Sixteen Gridded Precipitation Datasets over the Caribbean Region Using Gauge Observations},
JOURNAL = {Atmosphere},
VOLUME = {11},
YEAR = {2020},
NUMBER = {12},
ARTICLE-NUMBER = {1334},
URL = {https://www.mdpi.com/2073-4433/11/12/1334},
ISSN = {2073-4433},
DOI = {10.3390/atmos11121334}
}

@Article{Ahmed-2019,
AUTHOR = {Ahmed, K. and Sachindra, D. A. and Shahid, S. and Demirel, M. C. and Chung, E.-S.},
TITLE = {Selection of multi-model ensemble of general circulation models for the simulation of precipitation and maximum and minimum temperature based on spatial assessment
metrics},
JOURNAL = {Hydrology and Earth System Sciences},
VOLUME = {23},
YEAR = {2019},
NUMBER = {11},
PAGES = {4803--4824},
URL = {https://hess.copernicus.org/articles/23/4803/2019/},
DOI = {10.5194/hess-23-4803-2019}
}

@article{Katzav-2015,
	title = {The future of climate modeling},
	volume = {132},
	issn = {1573-1480},
	url = {https://doi.org/10.1007/s10584-015-1435-x},
	doi = {10.1007/s10584-015-1435-x},
	number = {4},
	journal = {Climatic Change},
	author = {Katzav, Joel and Parker, Wendy S.},
	month = oct,
	year = {2015},
	pages = {475--487}
}

@inbook{IPCC-2014, 
place={Cambridge}, 
title={Evaluation of Climate Models}, DOI={10.1017/CBO9781107415324.020}, 
booktitle={Climate Change 2013 – The Physical Science Basis: Working Group I Contribution to the Fifth Assessment Report of the Intergovernmental Panel on Climate Change}, 
publisher={Cambridge University Press}, author={Intergovernmental Panel on Climate Change}, year={2014}, 
pages={741–866}
}

@article {Nguyen-2017,
      author = "Phu Nguyen and Andrea Thorstensen and Soroosh Sorooshian and Qian Zhu and Hoang Tran and Hamed Ashouri and Chiyuan Miao and KuoLin Hsu and Xiaogang Gao",
      title = "Evaluation of CMIP5 Model Precipitation Using PERSIANN-CDR",
      journal = "Journal of Hydrometeorology",
      year = "2017",
      publisher = "American Meteorological Society",
      address = "Boston MA, USA",
      volume = "18",
      number = "9",
      doi = "10.1175/JHM-D-16-0201.1",
      pages= "2313 - 2330",
      url = "https://journals.ametsoc.org/view/journals/hydr/18/9/jhm-d-16-0201_1.xml"
}

@Article{RandomFields-2015,
    title = {Analysis, Simulation and Prediction of Multivariate Random Fields with Package {RandomFields}},
    author = {Martin Schlather and Alexander Malinowski and Peter J. Menck and Marco Oesting and Kirstin Strokorb},
    journal = {Journal of Statistical Software},
    year = {2015},
    volume = {63},
    number = {8},
    pages = {1--25},
    url = {http://www.jstatsoft.org/v63/i08/},
  }

@article{Swain-1991,
	title = {Color indexing},
	volume = {7},
	issn = {1573-1405},
	url = {https://doi.org/10.1007/BF00130487},
	doi = {10.1007/BF00130487},
	number = {1},
	journal = {International Journal of Computer Vision},
	author = {Swain, Michael J. and Ballard, Dana H.},
	month = nov,
	year = {1991},
	pages = {11--32}
}

@article{Almazroui-2021,
	title = {Projected {Changes} in {Temperature} and {Precipitation} {Over} the {United} {States}, {Central} {America}, and the {Caribbean} in {CMIP6} {GCMs}},
	volume = {5},
	issn = {2509-9434},
	url = {https://doi.org/10.1007/s41748-021-00199-5},
	doi = {10.1007/s41748-021-00199-5},
	number = {1},
	journal = {Earth Systems and Environment},
	author = {Almazroui, Mansour and Islam, M. Nazrul and Saeed, Fahad and Saeed, Sajjad and Ismail, Muhammad and Ehsan, Muhammad Azhar and Diallo, Ismaila and O’Brien, Enda and Ashfaq, Moetasim and Martínez-Castro, Daniel and Cavazos, Tereza and Cerezo-Mota, Ruth and Tippett, Michael K. and Gutowski, William J. and Alfaro, Eric J. and Hidalgo, Hugo G. and Vichot-Llano, Alejandro and Campbell, Jayaka D. and Kamil, Shahzad and Rashid, Irfan Ur and Sylla, Mouhamadou Bamba and Stephenson, Tannecia and Taylor, Michael and Barlow, Mathew},
	year = {2021},
	pages = {1--24}
}

@article{Bernton-2019,
    author = {Bernton, Espen and Jacob, Pierre E and Gerber, Mathieu and Robert, Christian P},
    title = "{On parameter estimation with the Wasserstein distance}",
    journal = {Information and Inference: A Journal of the IMA},
    volume = {8},
    number = {4},
    pages = {657-676},
    year = {2019},
    month = {10},
    issn = {2049-8772},
    doi = {10.1093/imaiai/iaz003},
    url = {https://doi.org/10.1093/imaiai/iaz003},
    eprint = {https://academic.oup.com/imaiai/article-pdf/8/4/657/31772180/iaz003.pdf}
}

@Article{Demirel-2018,
AUTHOR = {Demirel, M. C. and Mai, J. and Mendiguren, G. and Koch, J. and Samaniego, L. and Stisen, S.},
TITLE = {Combining satellite data and appropriate objective functions for improved spatial pattern performance of a distributed hydrologic model},
JOURNAL = {Hydrology and Earth System Sciences},
VOLUME = {22},
YEAR = {2018},
NUMBER = {2},
PAGES = {1299--1315},
URL = {https://hess.copernicus.org/articles/22/1299/2018/},
DOI = {10.5194/hess-22-1299-2018}
}

@Inbook{Villani-2009,
author="Villani, C{\'e}dric",
title="The Wasserstein distances",
bookTitle="Optimal Transport: Old and New",
year="2009",
publisher="Springer Berlin Heidelberg",
address="Berlin, Heidelberg",
pages="93--111",
isbn="978-3-540-71050-9",
doi="10.1007/978-3-540-71050-9_6",
url="https://doi.org/10.1007/978-3-540-71050-9_6"
}

@article {Tang-2021,
      author = "Guoqiang Tang and Martyn P. Clark and Simon Michael Papalexiou",
      title = "The Use of Serially Complete Station Data to Improve the Temporal Continuity of Gridded Precipitation and Temperature Estimates",
      journal = "Journal of Hydrometeorology",
      year = "2021",
      publisher = "American Meteorological Society",
      address = "Boston MA, USA",
      volume = "22",
      number = "6",
      doi = "https://doi.org/10.1175/JHM-D-20-0313.1",
      pages=      "1553 - 1568",
      url = "https://journals.ametsoc.org/view/journals/hydr/22/6/JHM-D-20-0313.1.xml"
}

@book{Shevlyakov-2016,
  title={Robust correlation: Theory and applications},
  author={Shevlyakov, Georgy L and Oja, Hannu},
  volume={3},
  year={2016},
  publisher={John Wiley \& Sons}
}

@article{Zamani-2019,
	title = {Evaluation of {CMIP5} models for west and southwest {Iran} using {TOPSIS}-based method},
	volume = {137},
	issn = {1434-4483},
	url = {https://doi.org/10.1007/s00704-018-2616-0},
	doi = {10.1007/s00704-018-2616-0},
	number = {1},
	journal = {Theoretical and Applied Climatology},
	author = {Zamani, Reza and Berndtsson, Ronny},
	month = jul,
	year = {2019},
	pages = {533--543},
}

@article{Brans-1986,
title = {How to select and how to rank projects: The Promethee method},
journal = {European Journal of Operational Research},
volume = {24},
number = {2},
pages = {228-238},
year = {1986},
note = {Mathematical Programming Multiple Criteria Decision Making},
issn = {0377-2217},
doi = {https://doi.org/10.1016/0377-2217(86)90044-5},
url = {https://www.sciencedirect.com/science/article/pii/0377221786900445},
author = {J.P. Brans and Ph. Vincke and B. Mareschal}
}

@article{Lai-1994,
title = {TOPSIS for MODM},
journal = {European Journal of Operational Research},
volume = {76},
number = {3},
pages = {486-500},
year = {1994},
note = {Facility Location Models for Distribution Planning},
issn = {0377-2217},
doi = {https://doi.org/10.1016/0377-2217(94)90282-8},
url = {https://www.sciencedirect.com/science/article/pii/0377221794902828},
author = {Young-Jou Lai and Ting-Yun Liu and Ching-Lai Hwang}
}

@article{Thakur-2022,
author = {Ritica Thakur and V. L. Manekar},
title = {Ranking of CMIP6 based High-resolution Global Climate Models for India using TOPSIS},
journal = {ISH Journal of Hydraulic Engineering},
volume = {0},
number = {0},
pages = {1-14},
year  = {2022},
publisher = {Taylor & Francis},
doi = {10.1080/09715010.2021.2015462},
URL = {https://doi.org/10.1080/09715010.2021.2015462},
eprint = {https://doi.org/10.1080/09715010.2021.2015462}
}

@article{Sithara-2022,
	title = {Statistical downscaling of sea levels: application of multi-criteria analysis for selection of global climate models},
	volume = {194},
	issn = {1573-2959},
	url = {https://doi.org/10.1007/s10661-022-10449-2},
	doi = {10.1007/s10661-022-10449-2},
	number = {10},
	journal = {Environmental Monitoring and Assessment},
	author = {Sithara, S. and Pramada, S. K. and Thampi, Santosh G.},
	month = sep,
	year = {2022},
	pages = {764},
}

@Article{Salabun-2020,
AUTHOR = {Salabun, Wojciech and Watróbski, Jaroslaw and Shekhovtsov, Andrii},
TITLE = {Are MCDA Methods Benchmarkable? A Comparative Study of TOPSIS, VIKOR, COPRAS, and PROMETHEE II Methods},
JOURNAL = {Symmetry},
VOLUME = {12},
YEAR = {2020},
NUMBER = {9},
ARTICLE-NUMBER = {1549},
URL = {https://www.mdpi.com/2073-8994/12/9/1549},
ISSN = {2073-8994}
}

@article{Maldonado-2018,
  title={A review of the main drivers and variability of Central America's Climate and seasonal forecast systems},
  author="Maldonado, Tito and Alfaro, Eric J. and Hidalgo, Hugo G.",
  journal={Revista de Biología Tropical},
  volume={66},
  number={1-1},
  pages={S153--S175},
  year={2018},
  doi = {https://doi.org/10.15517/rbt.v66i1.33294}
}

@article{Zhang2-2022,
  title={Evaluation of CMIP6 models toward dynamical downscaling over 14 CORDEX domains},
  author={Zhang, Meng-Zhuo and Xu, Zhongfeng and Han, Ying and Guo, Weidong},
  journal={Climate Dynamics},
  pages={1--15},
  year={2022},
  publisher={Springer},
  doi = {https://doi.org/10.1007/s00382-022-06355-5}
}

@misc{SMHI,
    author = "{Swedish Meteorological and Hydrological Institute}",
    title = {Why use a model ensemble?},
    year = {2016},
    url = {https://climateinformation.org/data-production-and-tailoring/why-use-a-model-ensemble/}
}

@misc{GFDL,
    author = "{Geophysical Fluid Dynamics Laboratory}",
    title = "{Climate Modeling}",
    url = {https://www.gfdl.noaa.gov/climate-modeling/},
    year = {2020}
}

\end{document}